\documentclass[letterpaper,11pt]{JHEP3}
\usepackage{amsmath}
\usepackage{amssymb}
\usepackage{bbm}
\usepackage{epsfig}
\usepackage{yfonts}
\newcommand{\labell}[1]{\label{#1}}

\newcommand{\be}{\begin{equation}}
\newcommand{\ee}{\end{equation}}
\newcommand{\bea}{\begin{eqnarray}}
\newcommand{\eea}{\end{eqnarray}}
\newcommand{\ba}{\begin{eqnarray}}
\newcommand{\ea}{\end{eqnarray}}

\newcommand{\beq}{\begin{equation}}
\newcommand{\eeq}{\end{equation}}
\newcommand{\beqa}{\begin{eqnarray}}
\newcommand{\eeqa}{\end{eqnarray}}
\newcommand{\beqar}{\begin{eqnarray*}}
\newcommand{\eeqar}{\end{eqnarray*}}

\newcommand{\reef}[1]{(\ref{#1})}

\newcommand{\eg}{{\it e.g.,}\ }
\newcommand{\ie}{{\it i.e.,}\ }

\newcommand{\mt}[1]{\textrm{\tiny #1}}

\newcommand{\veps}{\varepsilon}
\newcommand{\X}{\mathcal{X}}

\newcommand{\A}{\mathcal{A}}
\newcommand{\B}{\mathcal{B}}
\newcommand{\C}{\mathcal{C}}
\newcommand{\D}{\mathcal{D}}
\newcommand{\E}{\mathcal{E}}

\newcommand{\nvec}{{\vec{n}}}
\newcommand{\lgb}{\lambda_{\text{\tiny{GB}}}}
\newcommand{\tL}{\tilde{L}}

\newcommand{\lp}{\ell_{\mt P}}
\newcommand{\mbf}{\mathbf}
\newcommand{\fin}{f_\infty}
\newcommand{\ff}{f_{\infty}}
\newcommand{\ct}{C_{T}} 
\newcommand{\hi}{{\hat \imath}}
\newcommand{\hj}{{\hat \jmath}}

\newcommand{\nw}{N_W}

\preprint{arXiv:0911.4257 [hep-th]\\ UWO-TH-09/16}

\title{Holographic GB gravity in arbitrary dimensions}

\author{Alex Buchel,$^{a,b}$ Jorge Escobedo,$^{a,c}$ Robert C. Myers,$^{a}$ Miguel F. Paulos,$^{d}\qquad\qquad\qquad$
Aninda Sinha$^a$ and Michael Smolkin$^{a,e}$\\
$^a$ {\it Perimeter Institute for Theoretical Physics, Waterloo,
Ontario N2L 2Y5, Canada}\\
$^b$ {\it Department of Applied Mathematics, University of Western}\\ \ \ {\it Ontario, London, Ontario N6A 5B7, Canada}\\
$^c$ {\it Department of Physics and Astronomy and Guelph-Waterloo Physics Institute,}\\
\ \ {\it University of Waterloo, Waterloo, Ontario N2L 3G1, Canada}\\
$^d$ {\it Department of Applied Mathematics and Theoretical
    Physics, Cambridge CB3 0WA, UK}\\
$^e$ {\it Racah Institute of Physics, Hebrew University
Jerusalem 91904, Israel}}

\abstract{
We study the properties of the holographic CFT dual to
Gauss-Bonnet gravity in general $D\,(\ge 5)$ dimensions. We establish
the AdS/CFT dictionary and in particular relate the couplings of the
gravitational theory to the universal couplings arising in correlators
of the stress tensor of the dual CFT. This allows us to examine
constraints on the gravitational couplings by demanding consistency
of the CFT. In particular, one can demand positive energy fluxes in
scattering processes or the causal propagation of fluctuations.
We also examine the holographic hydrodynamics, commenting on the
shear viscosity as well as the relaxation time. The latter allows
us to consider causality constraints arising from the second-order
truncated theory of hydrodynamics.}

\begin{document}

\section{Introduction}

The AdS/CFT correspondence \cite{Maldacena, Witten, Gubser} or more
generally gauge/gravity dualities provide a theoretical framework in
which to study (certain) strongly coupled gauge theories. For example,
this approach allows for the calculation of transport coefficients for
gauge theory plasmas at strong coupling by means of relatively simple
supergravity computations, while these calculations are prohibitively
complicated by any other conventional methods --- for example, see
\cite{SonRomat,Rangamani,SonStarinets}. Recently this topic has been of
great interest motivated by the discovery of the strongly coupled
quark-gluon plasma (sQGP). In particular, it was found that the ratio
of shear viscosity to density entropy of any fluid dual to Einstein
gravity is precisely $1/4\pi$ \cite{KSS,einstein}\footnote{Recently this
universality has been extended to $T=0$ which is described
by extremal black holes in the bulk \cite{extr}.}. Despite the fact
that these holographic calculations deal with gauge theories which
are quite exotic compared to QCD, this result still seems to come
remarkably close to the value measured for the sQGP \cite{rick}.

Originally it was conjectured that these holographic calculations
provided a universal bound: $\eta/s\ge 1/4\pi$ \cite{KSS}. However, it
is now accepted that this conjectured bound is violated in string
theory by the effect of higher curvature interactions in the
gravitational action. Still the precise string theory constructions
where these higher curvature terms are under control only allow for
perturbative violations of the bound \cite{Kats,AlexRobAninda,
RobAnindaMiguel}. It is certainly also of interest to explore
situations where finite violations of the bound occur. A useful
framework for such explorations was found to be Gauss-Bonnet (GB)
gravity \cite{RobShenker, RobShenker2}. The original studies were made
with five-dimensional GB gravity theory which is dual to a
four-dimensional CFT but this analysis can easily be extended to any
$D\ge5$. Some work in this direction already appears in \cite{Sin,Jan}.
The present paper provides a comprehensive study of holographic GB
gravity in arbitrary dimensions.

An overview of the paper is as follows: We begin with a brief review of
Gauss-Bonnet (GB) gravity coupled to a negative cosmological constant
in section \ref{GBgrav}. In section \ref{dictionary}, we investigate
the AdS/CFT dictionary for these gravitational theories in an arbitrary
number of dimensions. In particular, we calculate the central charge
$\ct$ appearing in the two-point function of the stress tensor and the
parameters $t_2$ and $t_4$ appearing in the energy one-point function
describing certain scattering experiments, first proposed in \cite{HM}.
With these results, we determine the constraints on the GB coupling
arising from the requirement that the energy flux in the experiments is
everywhere positive. In section \ref{perturb}, we construct the
equations of motion for metric perturbations propagating in a black
hole background. These equations are then examined in section
\ref{causality} to study causality violations in the dual CFT. We find
that the constraints imposed on the GB coupling to avoid such
acausality precisely match the positive energy flux constraints derived
in section \ref{dictionary}. We examine holographic hydrodynamics for
GB gravity in section \ref{gohydro}. In particular, by studying the
propagation of sound waves in the dual plasma, we derive the relaxation
time, as well as the ratio of the shear viscosity to entropy density. Here we
also consider causality constraints within the framework of
second-order hydrodynamics. We conclude with a brief discussion of our
results in section \ref{discuss}. We also have some appendices
containing various technical details. In particular, appendix
\ref{misha} describes the calculation relating the scattering
parameters $t_2$ and $t_4$ to the couplings $\A$, $\B$ and $\C$ which
determine the three-point function of the stress-energy tensor.

While we were in the final stages of preparing this paper,
ref.~\cite{new} appeared which also explores causality constraints in
Gauss-Bonnet gravity.

\section{Gauss-Bonnet gravity} \label{GBgrav}

Consider GB gravity in $D\ge5$ spacetime dimensions, defined by the
following action
\begin{equation}
{I}_{\text{\tiny{GB}}}=\frac {1}{2\lp^{D-2}}\int d^Dx \sqrt{-g}
\left[\frac{(D-1)(D-2)}{L^2}+R+\frac{L^2\,\lgb}{(D-3)(D-4)}\X_4\right] \,,
\labell{SGB}
\end{equation}
where
\begin{equation}
 \X_4=R_{abcd}R^{abcd}-4 R_{ab}R^{ab}+R^2\ .
 \labell{euler4}
\end{equation}
Of course, this curvature-squared interaction is precisely the Euler
density of four-dimensional manifolds and so it does not effect the
gravitational equations of motion unless $D\ge5$. The solutions
describing planar AdS black holes take the form \cite{Cai}
\begin{equation}
ds^2=\frac{r^2}{L^2}\left(-\frac{f(r)}{\fin}
 \, dt^2 + \sum_{i=1}^{D-2} (dx^i)^2 \right) + \frac{L^2}{r^2} \,\frac{dr^2}{f(r)} \,,
\label{GBBH}
\end{equation}
where $f(r)$ is given by
 \be
f_\pm(r)=\frac{1}{2\lgb}\left[1\pm\sqrt{1-4\lgb\left(1-\frac{r_+^{D-1}}{r^{D-1}}
\right)}\ \right]\ .\labell{f}
 \eeq
In fact, in the following, we will only consider the solutions
$f=f_-(r)$ as these will be the only ones to correspond to nonsingular
black holes in a ghost-free vacuum \cite{GBghost,quasi}. Note that in
this class of solutions, the horizon appears at $r=r_+$. Using the
definition
 \beq
\fin=\lim_{r\rightarrow\infty}f(r)=\frac{1-\sqrt{1-4\lgb}}{2\,\lgb}\,,
 \labell{fine}
 \eeq
we have normalized the coordinates above so that
$\lim_{r\rightarrow\infty} g_{tt}/g_{xx}=-1$. This choice was made to
set the speed of light to one in the boundary metric (\ie in the dual
CFT).  Further, by setting $r_+ = 0$, we recover the AdS vacuum metric
in Poincar\'e coordinates. Examining $g_{rr}$, we can see that the AdS
curvature scale $\tL$ is related to the parameter $L$ in the action as
$\tL^2=L^2/\fin$.  In the following, we restrict our discussion to
$\lambda_{\text{\tiny{GB}}} < 1/4$ since no AdS vacua exist for larger
values of $\lgb$ as can be seen, \eg from eq.~\reef{fine}.

The Hawking temperature of this black hole solution is given by
 \beq
T=\frac{r_+}{4\pi L^2}\frac{D-1}{\sqrt{\fin}}\ . \labell{planarTh}
 \eeq
The energy and entropy densities are simply calculated as
 \beqa
\varepsilon &=& \frac{D-2}{2\sqrt{\fin}} \frac{r_+^{D-1}}{\lp^{D-2}
L^D}=2\pi \frac{D-2}{D-1}\left( \frac{4\pi\,{\fin}}{D-1}\right)^{D-2}\,
\left(\frac{\tL}{\lp}\right)^{D-2}
\,T^{D-1}\ , \labell{eden} \\
s&=&\frac{2\pi}{\lp^{D-2}}\left(\frac{r_+}{L}\right)^{D-2} =2\pi \left(
\frac{4\pi\,{\fin}}{D-1}\right)^{D-2}\,
\left(\frac{\tL}{\lp}\right)^{D-2}\,T^{D-2} \ .\labell{entropy}
 \eea
Further note that we find that $\varepsilon\propto T^d$ and $s\propto
T^{d-1}$ as expected for a CFT in $d=D-1$ dimensions (in the absence of
a chemical potential). Further, these expressions satisfy the precise
relation $\veps=\frac{d-1}{d} T s$, again as expected for a conformal
plasma.

\section{AdS/CFT dictionary} \label{dictionary}

In this section, we develop the dictionary relating the couplings in GB
gravity theory \reef{SGB} to parameters which characterize the dual
CFT. Since we are only dealing with the gravitational sector of the AdS
theory, we are looking to examine the behaviour of the stress energy
tensor of the CFT. So for example, the central charges, $c$ and $a$,
appearing in the trace anomaly are universal parameters characterizing
a four-dimensional CFT \cite{bd} and can be calculated in holographic
context \cite{renorm1}. For GB gravity, these calculations were
performed for $d=4$ in \cite{qthydro} and for $d=6$ in \cite{Jan}.
However, while the trace anomaly calculations can be extended to
examine CFT's in higher dimensions, the details change in each
dimension and the number of spacetime dimensions must be even. Hence we
do not pursue this approach here.

However, there is a ``central charge" common to CFT's in any number of
dimensions, $d$. This is the coefficient characterizing the leading
singularity in the two-point function of two stress tensors
\cite{EO,OP}:\footnote{Here and throughout the following, we are
assuming a Minkowski signature for the metric. Note that in
eq.~\reef{tensor}, $x_a=\eta_{ab}x^b$ (\ie $x_0=-t$).}
 \beq
\langle\, T_{ab}(x)\, T_{cd}(0)\, \rangle =\frac{\ct}{x^{2d}}\
{\mathcal I}_{ab,cd}(x)\,,
 \labell{twopt}
 \eeq
where
 \beq
 {\mathcal I}_{ab,cd}(x)=\frac{1}{2}
\left( I_{ac}(x)I_{bd}(x)+I_{ad}(x)I_{bc}(x)\right) -{1\over
d}\eta_{ab}\,\eta_{cd}\,,\labell{tensor0}
 \eeq
and
 \beq
I_{ab}(x)=\eta_{ab}-2\frac{x_a\,x_b}{x^2}\,.
 \labell{tensor}
 \eeq
This structure is completely dictated by the constraints imposed by
conformal symmetry and energy conservation \cite{EO,OP}. Of course, in
four dimensions, this coefficient is related to the standard central
charge $c$ which appears as the coefficient of the (Weyl)$^2$ term in
the trace anomaly: $\ct=(40/\pi^4)\,c$. Hence $\ct$ will be one of the
coefficients which we calculate holographically in the following to
establish our AdS/CFT dictionary for GB gravity in general dimensions.

To extend the dictionary further we need to identify additional
parameters that play an analogous universal role for CFT's in any
number of spacetime dimensions. With our focus on the stress energy
tensor, the obvious next step is to look for universal parameters in
the three-point function, as was extensively studied for CFT's in
general dimensions by \cite{EO,OP}. There it was shown that conformal
symmetry was powerful enough to determine the form of the three-point
function up to five constants, which are labeled $\A$, $\B$, $\C$, $\D$
and $\E$ in \cite{EO}. Conservation of the stress-energy imposes
further constraints which allow us to reduce the number of independent
parameters to three with:\footnote{They also find that this general
framework meets exceptions in low dimensions, with two independent
parameters in $d=3$ and one, in $d=2$. These reductions arise because
various tensor structures that are independent for $d\ge4$ are not
independent with a small number of dimensions.}
 \beqa
 \D&=&\frac{d^2-4}{2}\,\A+\frac{d+2}{2}\,\B-2d\,\C\,,
 \nonumber\\
 \E&=&(d^2-4)\,\A+\frac{d(d+6)}{4}\,\B-\frac{d(d+10)}{2}\,\C\,.
 \labell{elim}
 \eeqa
Further one finds that Ward identities relate the two- and three-point
functions and so $\ct$ can be expressed in terms of the parameters
characterizing the three-point function \cite{EO,OP}:
 \beq
\ct=\frac{\Omega_{d-1}}{2}\,
\frac{(d-1)(d+2)\,\A-2\,\B-4(d+1)\,\C}{d(d+2)}\,,
 \labell{central}
 \eeq
where $\Omega_{d-1}=2\pi^{d/2}/\Gamma(d/2)$ is the area of a unit
($d$--1)-sphere. In fact, one can perform a holographic calculation of
the three-point function \cite{rush}, however, extending these
calculations to GB gravity would require an exhaustive and exhausting
analysis. Therefore, we choose an indirect route to determining these
coefficients in the following.

In particular, we will consider extending the analysis of the energy
flux or ``energy one-point functions'' in \cite{HM} for CFT's in an
arbitrary spacetime dimension $d$.  This approach is to consider an
``experiment'' in which the energy flux was measured at null infinity
after a local disturbance was created by the insertion of the stress
tensor $\epsilon_{ij}\,T^{ij}$. The energy flux escaping at null
infinity in the direction indicated by the unit vector $\vec{n}$ then
takes the form
 \beq
\langle \E(\nvec)\rangle = \frac{E}{\Omega_{d-2}}\left[1 + t_2\,
\left(\frac{\veps^*_{ij}\veps_{i\ell}\, n^j
n^\ell}{\veps^*_{ij}\veps_{ij}} - \frac{1}{d-1}\right) + t_4\,
\left(\frac{|\veps_{ij}\,n^i
n^j|^2}{\veps^*_{ij}\veps_{ij}}-\frac{2}{d^2-1}\right)\right]\,,
 \labell{basic}
 \eeq
where $E$ is the total energy. The structure of this expression is
completely dictated by the symmetry of the construction.  Hence two
coefficients, $t_2$ and $t_4$, are constant parameters that
characterize the underlying CFT.\footnote{One comment is that this
general discussion only applies for $d\ge4$. It should perhaps be
evident that $d=3$ is an exception since in this case, there are not
enough spatial directions to consider rotations in the space orthogonal
to $\nvec$. In fact, we find that the starting point \reef{basic} is
not quite correct since
 \beq
\frac{\veps^*_{ij}\veps_{i\ell}\, n^j
n^\ell}{\veps^*_{ij}\veps_{ij}}=\frac{1}{2} \quad {\rm for}\ d=3\,.
 \labell{except}
 \eeq
Hence the most general expression for $d=3$ is simply
 \beq
\langle \E(\nvec)\rangle_{d=3} = \frac{E}{2\pi}\left[1 + t_4\,
\left(\frac{|\veps_{ij}\,n^i
n^j|^2}{\veps^*_{ij}\veps_{ij}}-\frac{1}{8}\right)\right]\,.
 \labell{basic1}
 \eeq
 \label{threed}}
Note that the (negative) constants appearing in the two factors
multiplied by $t_2$ and $t_4$ were chosen so that  these factors
contribute zero net flux when integrated over all directions. The
negative sign of these constants leads to interesting constraints on
the coefficients $t_2$ and $t_4$, which we discuss below in section
\ref{constraints}.

As presented in \cite{qthydro}, $t_2$ and $t_4$ can straightforwardly
be determined by a holographic calculation and we generalize these
calculations to GB gravity in any number of dimensions in section
\ref{t2t4} below. As discussed in \cite{HM}, the energy flux is
directly related to the three-point function and hence the two
coefficients $t_2$ and $t_4$ can be determined in terms of $\A$, $\B$
and $C$. A lengthy calculation presented in appendix \ref{misha} yields
\begin{eqnarray}
 t_2&=&\frac{2(d+1)}{d}\,{(d-2)\,(d+2)\,(d+1)\,\A+3\,d^{\,2}\,\B-4\,d\,(2d+1)\,\C
 \over (d-1)(d+2)\,\A-2\,\B-4(d+1)\,\C} ~,
\labell{misha2} \\ 
t_4&=&
-{(d+1) \over d}~{~(d + 2)\,(2d^2-3d-3)\,\A+2\,d^{\,2}\,(d+2)\,\B-4d\,(d+1)
\,(d+2)\,\C \over (d-1)(d+2)\,\A-2\,\B-4(d+1)\,\C}
 ~.\nonumber
\end{eqnarray}

As further discussed in \cite{HM}, a nonvanishing $t_4$ in a
four-dimensional CFT implies the action of the dual gravity theory must
contain terms cubic in the Riemann tensor. This analysis readily
extends to any number of dimensions and so since such interactions do
not appear in the GB gravity action \reef{SGB} studied here, the
holographic CFT must have $t_4=0$. Combined with eq.~\reef{misha2}, the
vanishing of $t_4$ imposes the constraint
 \beq
(d + 2)\,(2d^2-3d-3)\,\A+2\,d^{\,2}\,(d+2)\,\B-4d\,(d+1) \,(d+2)\,\C =
0
 \labell{misha3}
 \eeq
for the theories studied here.

\subsection{Central charge $\ct$} \label{CT}

First, we perform a holographic calculation of the central charge $\ct$
appearing in eq.~\reef{twopt}. We follow closely the derivation of the
two-point function given in \cite{hong}. We consider metric
fluctuations propagating in the AdS$_{d+1}$ vacuum geometry. It is
convenient to write the latter as $ds^2=\tilde L^2/u^2
(\eta_{ab}\,dx^adx^b +du^2)$ where as above
$\tL^2=L^2/\fin$.\footnote{Note that $u=(L\tilde L)/r$ where $r$ is the
radial coordinate used in section \ref{GBgrav}.} We choose a gauge
where the perturbation components $\delta g_{uu}=\delta g_{au}=0$ at
the AdS boundary. If we write the remaining components as $\delta
g_{ab}=\tL^2/u^2\,H_{ab}$, then the (on-shell) quadratic action for
$H_{ab}$ reduces to the following boundary term
 \be
I_2=\frac{\tilde L^{d-1}}{8\lp^{d-1}} (1-2\lgb\fin)\int_{\partial M}
d^d x\, u^{1-d} \, H_{ab}\, \partial_u H_{ab}\,,\labell{quad1}
 \ee
where the indices are simply contracted with $\delta_{ab}$. Imposing
the boundary conditions
 \be
H_{ab} (u=0,{\bf x})=\hat H_{ab}({\bf x})\,,\labell{boundc1}
 \ee
the full bulk solution for $H_{ab}$ can be written as
 \be
H_{ab}(u,{\bf x})=\frac{\Gamma[d]}{\pi^{d/2}
\Gamma[d/2]}\,\frac{d+1}{d-1} \int d^d x'\, \frac{u^d}{(u^2+|{\bf
x}-{\bf x'}|^2)^d}\, {\mathcal I}_{ab,cd}({\bf x}-{\bf x}')\, \hat
H_{cd}({\bf x'})\,,\labell{full1}
 \ee
where ${\mathcal I}_{ab,cd}$ precisely matches the tensor structure
appearing in eq.~\reef{tensor0}. Note that $\delta g_{uu}$ and $\delta
g_{au}$ are also nonvanishing in the bulk \cite{hong} but we ignore
these polarizations because they do not contribute in the quadratic
action \reef{quad1}. The quadratic action now becomes
 \be
I_2=(1-2\fin\lgb)\frac{\Gamma[d+1]}{\pi^{d/2}\Gamma[d/2]} \frac{\tilde
L^{d-1}}{8\lp^{d-1}} \frac{d+1}{d-1}\int d^d x\, d^d y\, \frac{\hat
H_{ab}({\bf x})\,{\mathcal I}_{ab,cd}({\bf x}-{\bf y})\,\hat
H_{cd}({\bf y})}{|{\bf x}-{\bf y}|^{2d}}\,.\labell{quad2}
 \ee
Varying the above expression with respect to $\hat H_{ab}$ then yields
the two-point function of the dual stress tensor and upon comparing
with (\ref{twopt}), we find
 \be
\ct=
\frac{d+1}{d-1}\frac{\Gamma[d+1]}{\pi^{d/2}\Gamma[d/2]}\,\frac{\tL^{d-1}}{\lp^{d-1}}
(1-2\fin\lgb)\,.\labell{ctfinal}
 \ee

\subsection{Holographic calculation of $t_2$ and $t_4$} \label{t2t4}

In this section, we perform a holographic computation of the energy
flux \reef{basic}. We follow closely the approach presented in
\cite{qthydro} and so only sketch the salient steps of the calculation.

We are interested in determining the energy flux
\be \langle \mathcal E(\mbf n)\rangle= \frac{ \langle 0| \mathcal
O^\dagger\, \mathcal E(\mbf n)\, \mathcal O|0\rangle}{\langle 0|
\mathcal O^\dagger\, \mathcal O|0 \rangle}\,. \labell{wok}\ee
In the present case, the state is being created by a tensor insertion
with $\mathcal O \sim \epsilon_{ij}T^{ij}(x)$ where the polarization
carries only spatial indices and $\epsilon_{ij}(x)\propto e^{-i E t}$.
Hence we see that this flux is determined by the three- and two-point
functions of the stress tensor in the CFT. This expression will depend
on the polarization tensor $\epsilon_{ij}$ and the unit vector $n^i$
indicating the direction in which the flux is measured. As noted above,
the $SO(d-1)$ invariance of the constructions fixes the final result to
take the form given in eq.~\reef{basic}.

If we adopt coordinates $x^\pm=x^0\pm x^{d-1}$, the energy flux
measured at future null infinity in direction specified by $n^i$ can be
written as
 \be
\mathcal E(\mbf n)=\lim_{x^+\to +\infty} \int_{-\infty}^{+\infty}dx^-\,
(x^+)^2\,\left(1+\frac{x_\hi x^\hi}{(x^+)^2}\right)\bigg (
T_{+j}^x(x^+,x^-,\mbf n)+T_{-j}^x(x^+,x^-,\mbf n)\bigg ) n^j\,.
\labell{flux1}
 \ee
We have introduced index notation where $\hi=1,\ldots,d-2$ while
$i,j=1,\ldots,d-1$ as usual. The superscript $x$ on the stress tensor
simply indicates that these operators are defined in flat space with
coordinates $x^a$. This notation is useful because next we introduce
new coordinates $y^a$ following \cite{HM}:
 \be
y^+=-\frac 1{x^+}\,, \quad y^-=x^--\frac{x_\hi x^\hi}{x^+}\,,\quad
y^{\hi}=\frac{x^{\hi}}{x^+}\,.
 \labell{ycoord}
 \ee
Note that the original Minkowski metric described by $x^a$ is conformal
to a flat space metric described by $y^a$:
 \be
ds^2=-dx^+ dx^-+(dx^\hi)^2=\frac{-dy^+ dy^-+(dy^\hi)^2}{(y^+)^2}\,.
 \labell{ymetric}
 \ee
The utility of transforming to $y^a$ is that null infinity,
$x^+\rightarrow\infty$, is now mapped to the (hyper)plane $y^+=0$. In
this plane, we also have
 \be
y^{\hi}=\frac{n^{\hi}}{1+n^{d-1}}\,.
 \labell{yplane}
 \ee
In terms of the new $y$ coordinates, the energy flux \reef{flux1}
becomes
 \be
\mathcal E(\nvec)=\Omega^{d-1} \,\int_{-\infty}^{+\infty} dy^- \,
T^y_{\ --}(y^+=0,y^-,y^\hi) \labell{operatorE}
 \ee
with $\Omega\equiv1/(1+n^{d-1})$. We have also introduced the $y$
superscript on the stress tensor to indicate the conformally
transformed operator:
 \be
T^y_{ab}=(y^+)^2\,\frac{\partial x^c}{\partial y^a} \, \frac{\partial
x^d}{\partial y^b}\,T^x_{cd}\,.
 \labell{ystress}
 \eeq

To compute $\langle \mathcal E(\mbf n)\rangle$ holographically, we must
turn on graviton perturbations whose boundary values source the
appropriate energy-momentum tensor insertions. As the first step
\cite{HM,hofman}, we consider the following shockwave
background\footnote{In this case, $u=L^4/r^2$ where $r$ is the radial
coordinate introduced in section \ref{GBgrav}.}
 \be
ds^2=\frac {L^2}{u}\bigg(\delta(y^+) W(u,y^\hi)(dy^+)^2-dy^+ dy^-+
(dy^\hi)^2\bigg)+\frac {L^2\, du^2}{4 \fin u^2}\,. \labell{shockw}
 \ee
In solving for the shockwave profile, the full nonlinear equations of
motion for GB gravity reduce to a simple linear equation for
$W(u,y^\hi)$ and in fact the curvature-squared terms do not contribute
\cite{hofman,qthydro}. The solution is chosen with the appropriate
asymptotic behaviour to source the operator $\mathcal E(n)\simeq \int
dy^- \,T_{--}$:
 \be
W(u,y^\hi)= \nw\, \Omega^{d-1}\,
\frac{u^{d/2}}{\bigg(u+(y^\hi-y'^\hi)^2\bigg)^{d-1}} \labell{solW}
 \ee
with $y'^\hi=n^{\hi}/(1+n^{d-1})$ as above and $\nw$ is a normalisation
constant.

It remains to add metric perturbations corresponding to the operators
$\mathcal O$ in eq.~\reef{wok}. For simplicity, we choose a
particular polarization with $\epsilon_{12}=a$ and all other components
vanishing. This description applies in the $x$ coordinate system and
the dual metric perturbation $\delta g_{ab}= L^2/u\, H_{ab}$ would have
the boundary condition
 \beq
H_{x^1x^2}(u=0) = a\, e^{-iEt}=a\, e^{-i \frac{E}2(x^++x^-)}
 \labell{boundH}
 \eeq
and all other components vanishing.  Next we wish to transform to the
$y$ coordinates and examine the overlap of the graviton with the
shockwave at $y^+=0$. Changing the coordinates, it can be shown
\cite{HM,qthydro} that the relevant graviton profile is given by
 \be
H_{y^1y^2}(y^+=0,y^-,y^\hi,u)\simeq \frac{1}{E^2}\, e^{-i E y^-/2}\
\delta^{d-2}(y^\hi)\,\delta(u-1)\,. \labell{tensorsol}
 \ee
In general, one also includes additional graviton polarizations to
ensure the perturbation is transverse and traceless in the bulk,
however, they will not contribute to the three-point function in GB
gravity \cite{qthydro}.

To find the three-point function, we add this perturbation to the
metric \reef{shockw} and evaluate the action on-shell. We are led to
examine terms proportional to $W (H_{y^1y^2})^2$. After integration by
parts and using the equations of motion, the relevant part of the cubic
effective action reduces to the following boundary integral:
 \begin{align}
I_3=-\frac {1}{8\lp^{d-1}}\int d^d y\,du\, \sqrt{-g}~H_{y^1y^2}\,
\partial_{-}^2H_{y^1y^2}\,W(u,y^\hi)\,\delta(y^+)\,\bigg(& 1-2 \ff \lgb\notag\\
&+
\frac{2\lgb\ff}{(D-3)(D-4)} T_2 \bigg)\,. \labell{threept}
 \end{align}
In the above expression,
 \be
T_2 \equiv \frac{\partial_1^2 W+\partial_2^2W-4\,
\partial_u W}{W}\,.\labell{threept2}
 \ee
After inserting the solution \reef{solW} for $W(u,y^\hi)$, we obtain
 \be
T_2=2(D-1)(D-2)\left (\frac{n_1^2+n_2^2}{2}-\frac 1{D-2}\right)\,.
 \labell{threept3}
 \ee
To normalize our result, we must divide by the two-point function
$\langle\, T_{12}\, T_{12}\,\rangle$, which was calculated in the
previous section. Upon fixing the normalisation $\nw$ of $W(u,y^\hi)$
appropriately, we find that the result takes the desired form
\reef{basic} with
 \be
\langle \mathcal
E(\theta)\rangle=\frac{E}{\Omega_{d-2}}\left[1+t_2\left(\frac{n_1^2+n_2^2}{2}-\frac
1{d-1}\right)\right]
 \ee
where the coefficient $t_2$ is given by
 \be
t_2=\frac{4 \ff \lgb}{1-2 \ff \lgb}\,\frac{(D-1)(D-2)}{(D-3)(D-4)}\,.
 \labell{t2holo}
 \ee
Implicitly, as expected, we have also found that $t_4=0$ for GB
gravity.

\subsection{Constraints} \label{constraints}

Because of the negative constants appearing in eq.~\reef{basic}, it is
easy to see that if the coefficients $t_2$ and $t_4$ become too large,
the energy flux measured in various directions will become negative.
Avoiding this problem then imposes various constraints, as in
\cite{HM}. To extract the various distinct constraints, we first fix
the unit vector, \eg $n^i=(1,0,0,\cdots)$. We then organize the
polarization tensors $\epsilon_{ij}$ according to their rotational
properties under the $SO(d-2)$ group that leaves $n^i$ invariant. There
are three possibilities and each produces a distinct constraint:
\begin{itemize}
\item Tensor (spin 2), \eg $\veps_{23}=\veps_{32}=a$ and all other
    components vanish,
 \beq
1-\frac{1}{d-1}\,t_2-\frac{2}{d^2-1}\,t_4\ge0
 \labell{cstrn1}
 \eeq
\item Vector (spin 1), \eg $\veps_{12}=\veps_{21}=a$ and all other
    components vanish,
 \beq
\left(1-\frac{1}{d-1}\,t_2-\frac{2}{d^2-1}\,t_4\right)
+\frac{1}{2}\,t_2\ge0
 \labell{cstrn2}
 \eeq
\item Scalar (spin 0), \eg $\veps_{ij}=a\times{\rm
    diag}(-(d-2),1,1,\cdots)$, \beq
\left(1-\frac{1}{d-1}\,t_2-\frac{2}{d^2-1}\,t_4\right)
+\frac{d-2}{d-1}\left(t_2+t_4\right)\ge0
 \labell{cstrn3}
 \eeq
\end{itemize}
These results provide a simple extension to a general dimension $d$ of
the constraints for four-dimensional CFT's derived in \cite{HM}.

Using the expressions in eq.~\reef{misha2}, the above constraints
\reef{cstrn1}, \reef{cstrn2} and \reef{cstrn3} can be translated to
constraints on the parameters $\A$, $\B$ and $\C$, \eg, using
eqs.~\reef{tenlin} and \reef{veclin}:
 \beqa
{\rm Tensor:}&& (d-2)\,(d+2)\,\A  + 2\,d\,\B - 4\,d\,\C\le0\,, \labell{cst1}\\
{\rm Vector:}&& (d-2)\,(d+2)\,\A+(3d-2)\,\B-8\,d\,\C\ge0\,, \labell{cst2}\\
{\rm Scalar:}&& \B-2\,\C\le0\,. \labell{cst3}
 \eeqa

Note that the three-point couplings obey an additional constraint
arising from the unitarity of the CFT. The latter implies that $\ct$ is
a positive quantity and so from eq.~\reef{central}, we have
 \beq
(d-1)(d+2)\,\A-2\,\B-4(d+1)\,\C>0\,.\labell{posit}
 \eeq
This inequality was already assumed in deriving the previous
constraints (\ref{cst1}--\ref{cst3}).

Before turning to the implications for the dual GB gravity, one might
consider the results for free fields \cite{OP,EO}\footnote{The general
results for the tensor fields are derived in Appendix
\ref{tensor-app}.}
 \beqa
\A&=&\ \ \frac{1}{\Omega_{d-1}^3}\left[\frac{d^3}{(d-1)^3}n_s
-\frac{d^3}{d-3}\tilde{n}_t\right]\,, \nonumber\\
\B&=&-\frac{1}{\Omega_{d-1}^3}\left[\frac{(d-2)d^3}{(d-1)^3}n_s
+\frac{d^2}{2}\tilde{n}_f+\frac{(d-2)d^3}{d-3}\tilde{n}_t
 \right]\,, \labell{freefield}\\
\C&=&-\frac{1}{\Omega_{d-1}^3}\left[\frac{(d-2)^2d^2}{4(d-1)^3}n_s
+\frac{d^2}{4}\tilde{n}_f+\frac{(d-2)d^3}{2(d-3)}\tilde{n}_t
 \right]\,, \nonumber
 \eeqa
where $\Omega_{d-1}$ is the area of a unit ($d$--1)-sphere, as defined
below eq.~\reef{central}. Also $n_s$ indicates the number of real
conformal scalars and $\tilde{n}_f$ is the number of (massless)
fermionic degrees of freedom. Hence $\tilde{n}_f=2^{\lfloor
d/2\rfloor}$ for a massless Dirac fermion in $d$ dimensions --- or in
the notation of \cite{EO}, $\tilde{n}_f={\rm tr}({\bf 1})$ where tr is
the Dirac trace. Finally for the case that $d=2n$, we have also
included the possible contribution of ($n$--1)-form potentials, for
which the standard free kinetic term is also conformally invariant.
Then $\tilde{n}_t$ denotes the number of degrees of freedom contributed
by these tensors. Generally, for a single ($n$--1)-form potential in
$d=2n$, we would have $\tilde{n}_t=\Gamma(2n-1)/\Gamma(n)^2$. For
example, with $d=4$, this would just be an Abelian vector field with
two degrees of freedom, \ie $\tilde{n}_t=2$. For $n$ odd, we might also
consider constraining the tensor by demanding that the field strength
be (anti-)self-dual, in which case the previous result for
$\tilde{n}_t$ would be multiplied by 1/2.

At this point, it is interesting to evaluate the constraints
(\ref{cst1}--\ref{cst3}) above with these free field results
\reef{freefield} for $\A$, $\B$ and $\C$
 \beqa {\rm Tensor:}&& (d-2)\,(d+2)\,\A  +
2\,d\,\B - 4\,d\,\C
=-\frac{(d^2-4)d^3}{d-3}\, \tilde{n}_t\,, \labell{cst1x}\\
{\rm Vector:}&& (d-2)\,(d+2)\,\A+(3d-2)\,\B-8\,d\,\C=\frac{1}{2}(d+2)
d^2\,\tilde{n}_f\,, \labell{cst2x}\\
{\rm Scalar:}&& \B-2\,\C=-\frac{(d^2-4)d^2}{2(d-1)^3}\,n_s\,.
\labell{cst3x}
 \eeqa
Hence with various combinations of the free fields, we are able to
precisely fill out the allowed region that is defined by requiring a
positive energy flux in eq.~\reef{basic}.\footnote{Momentarily, we are
treating $n_s$, $\tilde{n}_f$ and $\tilde{n}_t$ as continuous variables
here. This may be seen as a convenient approximation in the regime
where $\ct\gg1$. As a further note, recall that as discussed in
footnote \ref{threed}, the flux \reef{basic1} only contains a $t_4$
term in $d=3$. In this case, there are not enough spatial dimensions to
establish the tensor constraint \reef{cstrn1}. The vector and scalar
constraints reduce to
 \beq
 -4\le t_4\le4\,.
 \labell{cstr3}
 \eeq
With $d=3$, the free field expressions \reef{freefield} yield
$t_4=4(n_s-\tilde{n}_f)/(n_s+\tilde{n}_f)$ and hence the free theories
again fill the entire range of allowed couplings.} For example, with
$\tilde{n}_t=0$, we reach the boundary defined by the tensor constraint
in eq.~\reef{cst1}, \ie $(d-2)\,(d+2)\,\A  + 2\,d\,\B - 4\,d\,\C=0$.
This boundary surface is mapped out by allowing $n_s$ and $\tilde{n}_f$
to vary but note that eq.~\reef{cst2x} shows that $\tilde{n}_f=0$
corresponds precisely to the line where this tensor boundary intersects
that set by the vector constraint \reef{cst2}. Similarly, from
eq.~\reef{cst3x}, we see that $n_s=0$ corresponds to the intersection
of the boundaries set by the tensor \reef{cst1} and scalar \reef{cst3}
constraints.

Now using the AdS/CFT dictionary established above for GB gravity, we
may translate the constraints above to constraints on the GB coupling
$\lgb$. We begin with the constraint $\ct>0$ and comparing
eq.~\reef{ctfinal}, we find
 \beq
1-2\fin\lgb>0\,.\labell{posit2}
 \eeq
However, given the definition of $\fin$ in eq.~\reef{fine}, we find
that this constraint is always satisfied -- assuming $\lgb<1/4$,
otherwise no AdS vacua exist. This stems from choosing only to consider
the branch $f_-(r)$ in eq.~\reef{f}. In fact, the AdS vacua associated
with the `+' branch all fail to satisfy the above inequality. This
result can be directly related to the `ghost' behaviour of the
gravitons in the latter vacua, which is why we disgarded them in
section \ref{GBgrav}. Eq.~\reef{posit2} precisely matches the condition
that ensures that the graviton is not a ghost \cite{quasi}.

We now turn to the constraints in eqs.~(\ref{cstrn1}--\ref{cstrn3}).
Using the expression for $t_2$ in eq.~\reef{t2holo}, as well as $t_4=0$
and the definition \reef{fine} for $\fin$, we arrive at the following
constraints:
 \beqa
{\rm Tensor:}&& \lgb\le \frac{(D-3)(D-4)(D^2-3D+8)}{4(D^2-5D+10)^2}\,, \labell{cst1a}\\
{\rm Vector:}&& \lgb\ge-\frac{(D+1)(D-3)}{16}\,, \labell{cst2b}\\
{\rm Scalar:}&& \lgb\ge-\frac{(3D-1)(D-3)}{4(D+1)^2}\,. \labell{cst3c}
 \eeqa
The most stringent constraints here come from the tensor and scalar
channels which require
 \beq
-\frac{(3D-1)(D-3)}{4(D+1)^2}\le\lgb\le
\frac{(D-3)(D-4)(D^2-3D+8)}{4(D^2-5D+10)^2}\, \labell{range0}
 \eeq
in order that no negative energy fluxes appear in any channel. Similar
constraints on the GB coupling arise from demanding that the dual CFT
is causal \cite{RobShenker2}. For five-dimensional GB gravity, it was
found that these two sets of constraints were identical
\cite{HM,hofman,RobAlex}. Below we will show that the flux constraints
above again match the causality constraints for an arbitrary $D$.

\section{Gauss-Bonnet black hole perturbations} \label{perturb}

The topics of the two subsequent sections will be the study of
causality violations (section \ref{causality}) and hydrodynamics
(section \ref{gohydro}) in the CFT's dual to GB gravity. In both cases,
the analysis focuses on examining metric perturbations propagating in
the GB black hole background \eqref{GBBH}. Hence it is useful to
outline the general framework and to derive the equations of motion for
these perturbations in the present section. To begin, we review the
techniques introduced in \cite{KovtunStarinets} to study black hole
quasinormal modes. The specific background of interest is the GB black
hole \reef{GBBH} introduced above. For the following analysis, it is
convenient to choose the radial coordinate $u=r_+^2/r^2$, in which case
the metric becomes
\begin{equation}
ds^2=\frac{r_+^2}{u L^2}\left(-\frac{f(u)}{\fin} \, dt^2
+ \sum_{i=1}^{D-2} dx_i^2 \right) + \frac{L^2}{4u^2f(u)} du^2,
\label{GBBH2}
\end{equation}
with
\begin{equation}
f(u)=\frac{1-\sqrt{1-4\lgb (1-u^{\frac{D-1}{2}})}}{2\lgb}
\label{f2}
\end{equation}
With this radial coordinate, the horizon is now located at $u=1$ and
the boundary at $u=0$.

Now on this background \reef{GBBH2}, we wish to study gravitational
fluctuations $h_{\mu\nu}$. The latter are taken to be of the form
$h_{\mu\nu}=h_{\mu\nu}(u)e^{-i \omega t+iqz}$, where we choose the
direction $z$ to be the planar coordinate $x^{D-2}$.  As usual
according to the AdS/CFT dictionary, these perturbations are dual to
stress-energy probes of a finite temperature plasma in the boundary
CFT. For simplicity, we restrict to the case where the metric
perturbations do not couple to fluctuations of other background fields.
The gravitational fluctuations $h_{\mu \nu}$ can be classified
according to their transformation properties under the remaining
symmetry group $SO(D-3)$ acting in the $x^\hi$ directions --- where
$\hi = 1, \dots,{D-3}$, as in the previous section. Hence, $h_{tt}$,
$h_{tz}$, $h_{zz}$, $h_{uu}$, $h_{tu}$ and $h_{zu}$ transform trivially
under these rotations and can be considered as spin 0
perturbations.\footnote{Note that the present classification of the
probes is identical to that discussed for the states in the scattering
processes in section \ref{constraints}. Unfortunately here, we cannot
refer the spin 0 perturbations as scalar perturbations because this
nomenclature clashes with the hydrodynamic terminology -- see below.}
Similarly, $h_{t \hi}$, $h_{z \hi}$ and $h_{u \hi}$ transform as as
spin 1 perturbations and $h_{\hi \hat\jmath}$ transform as spin 2
tensors. Therefore, we have three symmetry channels for gravity
perturbations, conventionally referred to as the sound, shear and
scalar channels, where the terminology is adopted from the hydrodynamic
description of the dual CFT plasma. The equations of motion for
fluctuations belonging to different symmetry channels are guaranteed to
decouple due to the $SO(D-3)$ symmetry. Following
\cite{KovtunStarinets}, it is convenient to rescale the fluctuations as
\begin{equation}
H_{tt}=\frac{u L^2}{r_+^2}\frac{\fin}{f(u)}h_{tt}\,,
\quad H_{tz}=\frac{u L^2}{r_+^2}h_{tz}\,,
\quad H_{ij}=\frac{u L^2}{r_+^2}h_{ij}\,, \quad H=\sum_{\hi}H_{\hi \hi}
=\frac{u L^2}{r_+^2}\frac{h}{(D-3)}\,,
\labell{Hs}
\end{equation}
with $i,j \neq t$ and $h=\sum_{\hi} h_{\hi \hi}$. In terms of these, we
make a particular choice of perturbations for each of the symmetry
channels for further study in the following:
 \beqa
\text{Sound channel (spin 0):}&
\quad& H_{tt}, H_{tz}, H_{zz}, H_{uu}, H_{tu}, H_{zu}, H, \nonumber \\
\text{Shear channel (spin 1):}&
\quad& H_{t x}, H_{z x}, H_{u x},  \nonumber \\
\text{Scalar channel (spin 2):}& \quad& H_{x y}\,.
\labell{fluctuations2}
 \eeqa
Note that we have renamed $x^1=x$ and $x^2=y$ for notational
convenience.  Using the momentum and frequency of the fluctuation, we
introduce the following dimensionless quantities
\begin{equation}
\textswab{q}=\frac{q}{2\pi T}, \qquad \textswab{w}=\frac{\omega}{2\pi T},
\labell{qw}
\end{equation}
where $T$ is the Hawking temperature \reef{planarTh} of the black hole.
Furthermore, we can define the following ``gauge-invariant'' variables
in each of the channels
 \beqa
\text{Sound channel:}& \quad& Z_{\text{\tiny{sound}}} \equiv
\textswab{q}^2 \frac{f(u)}{\fin} \, H_{tt} + 2 \textswab{w} \,
\textswab{q} \, H_{tz} + \textswab{w}^2 H_{zz}  + \textswab{q}^2
\left(\frac{f(u)}{\fin}-\frac{u f'}{\fin}-\frac{\textswab{w}^2}{\textswab{q}^2}\right)
H, \nonumber \\
\text{Shear channel:}& \quad& Z_{\text{\tiny{shear}}} \equiv
\textswab{q}
\, H_{t x}+\textswab{w} \, H_{z x}, \nonumber \\
\text{Scalar channel:}& \quad& Z_{\text{\tiny{scalar}}} \equiv H_{x
y}\,. \labell{channels}
 \eeqa
They are gauge-invariant in the sense that they are invariant under the
residual
infinitesimal diffeomorphisms: $h_{\mu\nu} \to h_{\mu\nu} - \nabla_\mu
\xi_\nu- \nabla_\nu \xi_\mu$, where $\xi_{\mu}=\xi_{\mu}(r)e^{-i \omega
t+iqz}$  and the covariant derivatives are taken with respect to the
background metric \eqref{GBBH2}, which preserve the background metric \reef{GBBH2}. By first finding the equations of
motion obeyed by the rescaled fluctuations $H_{\mu \nu}$, we will be
able to derive second-order differential equations for each of the $Z$'s.

To obtain the equations of motion for $H_{\mu \nu}$, we perturb the
black hole metric \eqref{GBBH} by $ds^2_{pert}$, which contains an
infinitesimal parameter $\epsilon$. Then, we evaluate the GB Lagrangian
as a series in $\epsilon$ and pick up the coefficient of the
second-order term.  By varying this resulting Lagrangian with respect
to each of the fluctuations we obtain the equations of motion of
interest. We now present each of the three cases above in turn.

\subsection{Scalar channel}
The scalar channel is the simplest case as we only need to consider one
perturbation, namely
\begin{equation*}
ds^2_{\text{pert}}= 2\epsilon \, \frac{r_+^2}{u L^2} \left(H_{xy}\,dx\,dy\right)\,.
\end{equation*}
The  resulting equation of motion for the perturbation
$Z_{\text{\tiny{scalar}}} \equiv H_{xy}$ is the following:
\begin{equation*}
Z''_{\text{\tiny{scalar}}} + \mathcal{C}^{(1)}_{\text{\tiny{scalar}}}
\, Z'_{\text{\tiny{scalar}}} + \mathcal{C}^{(2)}_{\text{\tiny{scalar}}} \, Z_{\text{\tiny{scalar}}}=0
\,,
\end{equation*}
with the following expressions for the coefficients
\begin{equation*}
\mathcal{C}^{(1)}_{\text{\tiny{scalar}}}= \frac{\mathcal{P}}{u f \mathcal{N} \mathcal{M}}\,,
\end{equation*}
\begin{equation*}
\mathcal{C}^{(2)}_{\text{\tiny{scalar}}}= \frac{\mathcal{R}}{u f^2 \mathcal{N} \mathcal{M}}\,.
\end{equation*}
Here and in the following, the expressions denoted by capital
calligraphic letters are expressions that we define in Appendix
\ref{appA} in order to present our results as succinctly as possible.
These scalar channel equations were already obtained for GB gravity in
general spacetime dimensions by \cite{Sin}.

\subsection{Shear channel}
In the shear channel, we perturb the black hole with
\begin{equation*}
ds^2_{\text{pert}}= 2\epsilon \, \frac{r_+^2}{u L^2} \left(H_{tx}\,dt\,dx +H_{zx}\,dz\,dx + H_{ux}\,du\,dx\right).
\end{equation*}
We choose to work in the gauge where $H_{ux}=0$. The resulting
equations that we get from varying the action with respect to $H_{tx}$,
$H_{zx}$ and $H_{ux}$ are, respectively
\begin{equation*}
H''_{tx}-\frac{(D-3)}{2} \frac{\mathcal{M}}{u \, \mathcal{N}} \, H'_{tx} - \frac{(D-1)^2}{16\fin}\frac{ \textswab{q}\, \mathcal{M}}{u f \mathcal{N}} \left(\textswab{q}\,H_{tx}+\textswab{w}\,H_{zx}\right)=0,
\end{equation*}
\begin{equation*}
H''_{zx}+ \frac{\mathcal{P}}{u f \mathcal{N} \mathcal{M}} \, H'_{zx} + \frac{(D-1)^2}{16}\frac{\textswab{w}}{u f^2} \left(\textswab{q}\,H_{tx}+\textswab{w}\,H_{zx}\right)=0,
\end{equation*}
\begin{equation*}
H'_{zx}+\frac{\textswab{w} \fin \mathcal{N}}{ \textswab{q} f \mathcal{M}} \, H'_{tx}=0.
\end{equation*}
Next, we multiply the first equation by $\textswab{q}$ and the second
one by $\textswab{w}$, add both equations and, using the third
equation, find a differential equation for the gauge-invariant variable
$Z_{\text{\tiny{shear}}} \equiv \textswab{q} \, H_{tx}+\textswab{w} \,
H_{zx}$. The result is
\begin{equation}
Z''_{\text{\tiny{shear}}} + \mathcal{C}^{(1)}_{\text{\tiny{shear}}} \, Z'_{\text{\tiny{shear}}} + \mathcal{C}^{(2)}_{\text{\tiny{shear}}} \, Z_{\text{\tiny{shear}}}=0,
\label{sheareom}
\end{equation}
with the following expressions for the coefficients
\begin{equation}
\mathcal{C}^{(1)}_{\text{\tiny{shear}}}= \frac{-2\textswab{w}^2 \fin \mathcal{P}\mathcal{N}-(D-3) \textswab{q}^2 f^2 \mathcal{M}^3}{2 u f \mathcal{M}\mathcal{N}\left( \textswab{q}^2 f \mathcal{M} - \textswab{w}^2 \fin\mathcal{N} \right)},
\label{c1shear}
\end{equation}
\begin{equation}
\mathcal{C}^{(2)}_{\text{\tiny{shear}}}= \frac{(D-1)^2}{16\fin} \left(\frac{\textswab{w}^2\fin \mathcal{N} -  \textswab{q}^2 f \mathcal{M}}{u f^2 \mathcal{N}}\right).
\label{c2shear}
\end{equation}

\subsection{Sound channel}
Lastly, let us consider the following perturbation in the sound channel
\begin{align*}
ds^2_{\text{pert}}= \epsilon \, \frac{r_+^2}{u L^2}
\Bigl(&\frac{f}{\fin}\, H_{tt}\, dt^2 + 2H_{tz}\,dt\,dz +(D-3)H_{\hi \hi}\, (dx^\hi)^2
+H_{zz}\, dz^2 + \notag \\
&+2H_{tu}\,dt\,du + 2H_{zu}\,dz\,du + H_{uu}\,du^2 \Bigr)\,.
\end{align*}
Choosing the gauge $H_{tu}=H_{zu}=H_{uu}=0$, the equations of motion
that we obtain from varying the action with respect to $H_{tt}$,
$H_{tz}$, $H=\sum_{\alpha}H_{\alpha \alpha}$, $H_{zz}$, $H_{tu}$,
$H_{zu}$ and $H_{uu}$, respectively, are the following:
\begin{equation*}
H''+\frac{H''_{zz}}{(D-3)}+ \frac{\mathcal{T}}{u f \mathcal{N}} \, H' + \frac{\mathcal{T}}{(D-3)\,u f \mathcal{N}} \, H'_{zz} -\frac{(D-1)^2}{16\fin}\frac{ \textswab{q}^2 \mathcal{M}}{u f \mathcal{N}} \, H=0,
\end{equation*}
\begin{equation*}
H''_{tz}-\frac{(D-3)}{2} \frac{\mathcal{M}}{u \, \mathcal{N}} \, H'_{tz} +\frac{(D-1)^2(D-3)}{16\fin}\frac{ \textswab{q} \, \textswab{w} \, \mathcal{M}}{u f \mathcal{N}} \, H=0,
\end{equation*}
\begin{align*}
H''_{tt}&-(D-4)\frac{\mathcal{M}}{\mathcal{N}} \, H'' - \frac{\mathcal{M}}{\mathcal{N}} \, H''_{zz}+\frac{\mathcal{S}}{u f \mathcal{N}} \, H'_{tt} - (D-4) \frac{\mathcal{P}}{u f \mathcal{N}^2} \, H' + \notag \\
&- \frac{\mathcal{P}}{u f \mathcal{N}^2} \,H'_{zz} -\frac{(D-1)^2}{8} \frac{\textswab{q} \, \textswab{w} \,  \mathcal{M}}{u f^2 \mathcal{N}} \, H_{tz} -\frac{(D-1)^2}{16\fin} \frac{ \textswab{q}^2 \mathcal{M}}{u f \mathcal{N}} \, H_{tt} + \notag \\
&-(D-4)\frac{\mathcal{R}}{u f^2 \mathcal{N}^2} \, H -\frac{(D-1)^2}{16} \frac{\textswab{w}^2 \mathcal{M}}{u f^2 \mathcal{N}} \, H_{zz}=0,
\end{align*}
\begin{equation*}
H''_{tt}-(D-3)\frac{\mathcal{M}}{\mathcal{N}} \, H'' +\frac{\mathcal{S}}{u f \mathcal{N}} \, H'_{tt} - (D-3)\frac{\mathcal{P}}{u f \mathcal{N}^2}\,H' -\frac{(D-1)^2(D-3)}{16} \frac{\textswab{w}^2 \mathcal{M}}{u f^2 \mathcal{N}} \,H=0,
\end{equation*}
\begin{align*}
&H'+\frac{H'_{zz}}{(D-3)}+\frac{\textswab{q}}{(D-3)\textswab{w}} \, H'_{tz} + \frac{(D-1)}{2(D-3)}\frac{\textswab{q} \, \mathcal{K}}{u f \textswab{w} \, \mathcal{N}^{\frac{1}{2}}} \, H_{tz} + \frac{(D-1)}{4} \frac{\mathcal{K}}{u f \mathcal{N}^{\frac{1}{2}}} \,
H\notag\\
&+ \frac{(D-1)}{4(D-3)} \frac{\mathcal{K}}{u f \mathcal{N}^{\frac{1}{2}}} \, H_{zz}=0,
\end{align*}
\begin{equation*}
H'_{tz}+\frac{ \textswab{q} f}{\textswab{w} \fin} \, H'_{tt} -\frac{(D-3)\textswab{q} f \mathcal{M}}{\textswab{w} \fin \mathcal{N}}\, H' - \frac{(D-1)}{4\fin} \frac{ \textswab{q} \, \mathcal{K}}{u \, \textswab{w} \, \mathcal{N}^{\frac{1}{2}}} \, H_{tt}=0,
\end{equation*}
\begin{align*}
(D-3)\,H'&+H'_{zz} -\frac{f \mathcal{N}}{\mathcal{V}} \, H'_{tt} - \frac{(D-1)^2(D-3)}{8(D-2)\mathcal{V}}\left(\frac{\textswab{w}^2 \mathcal{N}}{f} - \frac{ \textswab{q}^2 \mathcal{M}}{\fin}\right) H - \frac{(D-1)^2}{8(D-2)} \frac{\textswab{w}^2 \mathcal{N}}{f \, \mathcal{V}} \, H_{zz}+ \notag \\
&- \frac{(D-1)^2}{8(D-2)} \frac{\textswab{q}^2 \mathcal{N}}{\fin\mathcal{V}} \, H_{tt} - \frac{(D-1)^2}{4(D-2)} \frac{\textswab{q} \, \textswab{w} \, \mathcal{N}}{f \, \mathcal{V}} \, H_{tz}=0.
\end{align*}

Again, the goal is to use these equations of motion to write a
second-order differential equation for the gauge-invariant variable
\begin{equation}
Z_{\text{\tiny{sound}}} \equiv  \frac{\textswab{q}^2 f}{\fin} \, H_{tt} + 2 \textswab{w} \, \textswab{q} \, H_{tz} + \textswab{w}^2 H_{zz}  + \frac{\textswab{q}^2 f}{\fin} \left(1+\frac{(D-1) \mathcal{K}}{2f \mathcal{N}^{1/2}}-\frac{\textswab{w}^2\fin}{ \textswab{q}^2 f}\right) H.
\end{equation}
After a series of algebraic manipulations, we arrive at the following equation
\begin{equation}
Z''_{\text{\tiny{sound}}} + \mathcal{C}^{(1)}_{\text{\tiny{sound}}} \, Z'_{\text{\tiny{sound}}} + \mathcal{C}^{(2)}_{\text{\tiny{sound}}} \, Z_{\text{\tiny{sound}}}=0,
\label{soundeom}
\end{equation}
where the expressions for the coefficients
$\mathcal{C}^{(1)}_{\text{\tiny{sound}}}$ and
$\mathcal{C}^{(2)}_{\text{\tiny{sound}}}$ are extremely long and are given
in Appendix \ref{appB}.

\section{Causality constraints} \label{causality}

Having obtained the equations of motion for the gauge-invariant
variables $Z$ in each of the channels, we now proceed to use them to
study causality violation in GB gravity, following
\cite{RobShenker,RobShenker2}. We will focus on the shear and sound
channels because the scalar channel was already considered for GB gravity
in arbitrary dimensions by \cite{Sin}.  Let us first outline the
general strategy behind this approach, which will then be used in each
of the channels separately. First, note that in each of the channels,
the differential equation for $Z$ has the form
\begin{equation}
Z''(u)+ \mathcal{C}^{(1)} Z'(u) +  \mathcal{C}^{(2)} Z(u)=0.
\label{eqZ}
\end{equation}
The general idea given in \cite{RobShenker2} consists of introducing a
new radial coordinate $\rho$ and rescaling the profile $Z=\psi(u)/B(u)$
to bring this equation \reef{eqZ} into the form of an effective
Schr\"{o}dinger equation,
\begin{equation}
- \partial^2_{\rho}\, \psi + U\, \psi=\textswab{w}^2 \psi\,.
\labell{schrod2}
\end{equation}
This form can be achieved with $\rho$ and $B$ defined by
 \beqa
\frac{d\rho}{du}&=&- \frac{(D-1)}{4\,u^{1/2} f(u)}\,, \labell{rho}\\
\frac{d \ln B}{du}&=&\frac{\mathcal{C}^{(1)}}{2} -
\frac{1}{4u}+\frac{(D-1)u^{\frac{D-1}{2}}}{4uf
\left(1-2\lambda_{\text{\tiny{GB}}} f\right)}\,.\labell{BBB}
 \eeqa
Further, we introduce $\hbar \equiv 1/\textswab{q}$ and $\alpha \equiv
\textswab{w}/\textswab{q}$. Then upon dividing the above equation
\reef{schrod2} by $\textswab{q}^2$ and taking the limit $\textswab{q}
\to \infty$ (or equivalently $\hbar \to 0$) with $\alpha$ fixed, we
arrive at
\begin{equation}
- \hbar^2 \partial^2_{\rho}\, \psi + \left(U^0 + \hbar^2 U^1 +\cdots \right) \psi=\alpha^2 \psi.
\label{schrod3}
\end{equation}
Note that the effective potential in this equation was obtained by
Taylor expanding $U$ in the limit $\hbar \to 0$.  Note that with the
above construction, the potential $U^0$ only depends on $u$, $D$ and
$\lambda_{\text{\tiny{GB}}}$, whereas $U^1$ and higher order terms may
also depend on $\alpha$. The key point is that in the $\hbar\to0$
limit, the dominant contribution to the effective potential comes from
$U^0$. Hence, for our purposes, it is sufficient to study the equation
\begin{equation}
- \hbar^2 \partial^2_{\rho}\, \psi + U^0\, \psi=\alpha^2 \psi\,.
\label{schrodfinal}
\end{equation}

All the preceding is applicable to any of the three channels that we
are considering. We will now study the behaviour of equation
\eqref{schrodfinal} in each of these channels and give arguments as to
what conditions have to be satisifed in order to preserve causality in
the dual CFT.

\subsection{Shear channel}
The leading potential (as $\hbar\to 0$) in the shear channel is
\begin{equation}
U^0_ {\text{\tiny{shear}}}=\frac{ \left[-3+D-2(D-1)
\lambda_{\text{\tiny{GB}}}-2(D-5)\lambda_{\text{\tiny{GB}}}
f \left(1-\lambda_{\text{\tiny{GB}}} f \right) \right] f }{(D-3)
(1-2\lambda_{\text{\tiny{GB}}} f)^2\fin}.
\labell{U0shear}
\end{equation}
while the expression for the subleading potential $U^1_
{\text{\tiny{shear}}}$ is too long to be presented here.  With this
explicit formula at hand, we can plot the leading potential as a
function of $u$ (using \eqref{f} and \eqref{fine}) in the physical
region $u \in [0,1]$.  For any $\lgb < 1/4$, we have that $U^0_
{\text{\tiny{shear}}}(u=0)=1$ and $U^0_ {\text{\tiny{shear}}}(u=1)=0$.
For small  values of $|\lgb|$, $U^0$ is a monotonically decreasing
function between the boundary and the horizon. However, as we make
$\lambda_{\text{\tiny{GB}}}$ more negative, the potential develops a
single maximum between $u=0$ and $u=1$.  This behaviour is illustrated
in figure \ref{figU0shear}.
\EPSFIGURE{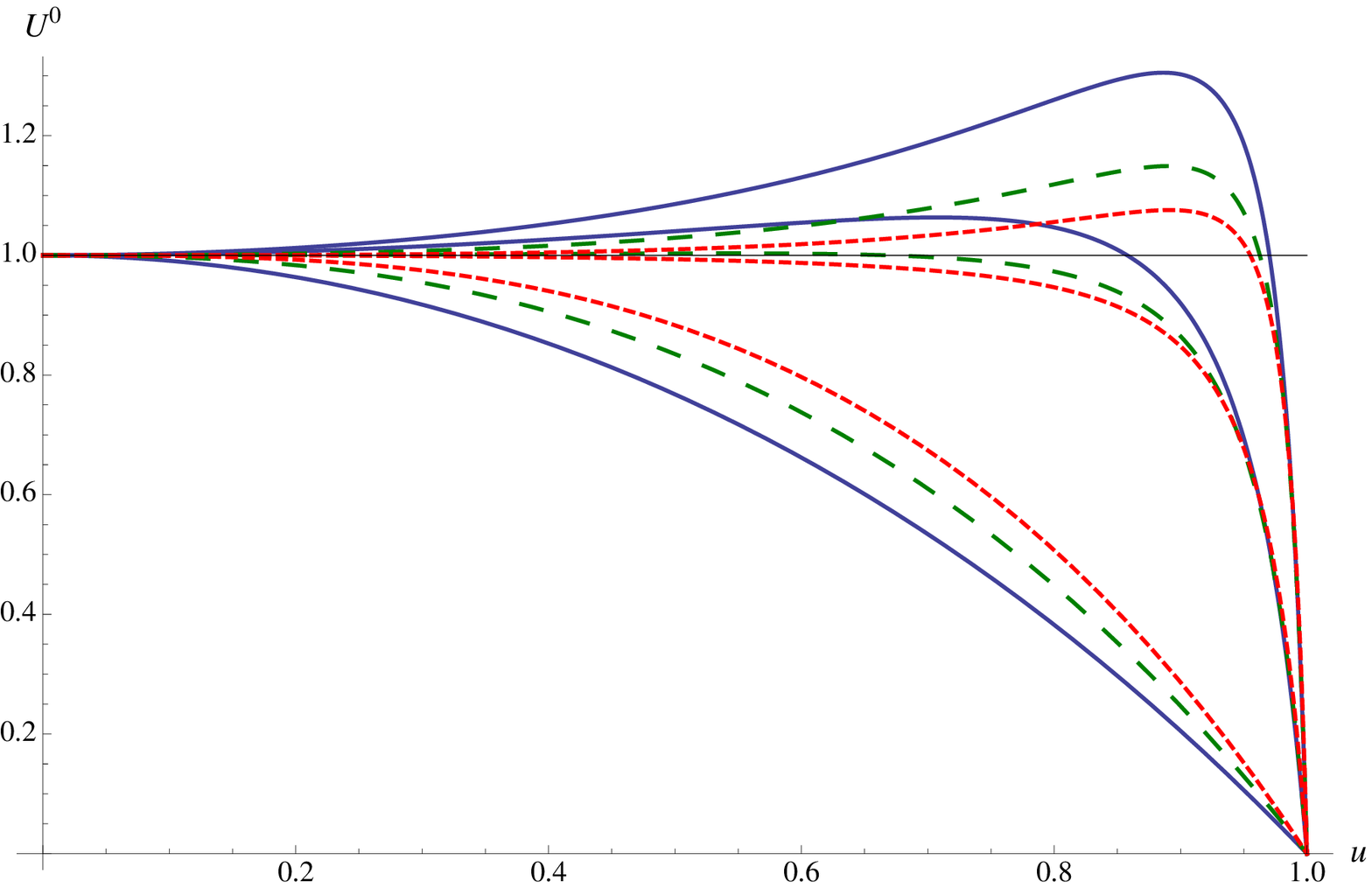,width=12cm} {(Colour online) Leading potential $U^0$ for the
Schr\"{o}dinger-like equation \reef{schrodfinal} in the shear channel.
The blue (solid), green (long dash) and red (small dash) lines
correspond to $D=5$, $6$ and $7$, respectively. For each dimension, the
three potentials correspond to $\lgb=-0.02,-1.5,-3.5$, from bottom to
top. We also show the line $U^0=1$. The behaviour for higher $D$ is
similar.\label{figU0shear}}

The appearance of a new maximum at some $0<u<1$ implies the existence
of quasinormal modes with $\mathop{\rm Re} (\alpha^2) \simeq >1$.  In
turn, this implies that in the limit $\textswab{q} \to \infty$, in
which we wrote the Schr\"{o}dinger equation, $\mathop{\rm Re}
(\textswab{w})/\textswab{q} > 1$ for these modes, leading to a
violation of causality in the dual CFT.  Hence to avoid causality
violation in this channel, we will impose a bound on $\lgb$ in order to
avoid the appearance of a maximum in the potential.

It is clear from the plot in figure \ref{figU0shear} that there is some
critical value $\lambda_{\text{\tiny{GB}}}^ {\text{\tiny{shear}}}$
below which $U^0$ exhibits a maximum.  To determine the precise value,
we first Taylor expand the potential around the boundary $u=0$
obtaining
\begin{equation}
U^0_ {\text{\tiny{shear}}}=1-\frac{(D-3)\left(
1+\sqrt{1-4\lambda_{\text{\tiny{GB}}}}\right) + 8\lambda_{\text{\tiny{GB}}}}{2(D-3)
\left(1-4\lambda_{\text{\tiny{GB}}}\right)} \, u^{\frac{D-1}{2}} + \mathcal{O}\left(
u^{D-1}\right)\,.
\labell{U0shearexp}
\end{equation}
The sign of the coefficient of the $u^{(D-1)/{2}}$ term will determine
whether or not a maximum appears in $U^0$ at $u>0$. The vanishing of
this coefficient then determines the critical value of the GB coupling:
\begin{equation}
\lambda_{\text{\tiny{GB}}}^ {\text{\tiny{shear}}}= - \frac{(D+1)(D-3)}{16}\,.
\labell{lgbshear}
\end{equation}
If the GB coupling is greater than or equal to this value, the
potential monotonically decreases to $0$ at $u=1$.  Therefore, we
obtain the following lower bound on $\lgb$ which has to be
satisfied to preserve causality in the shear channel:
\begin{equation}
\lambda_{\text{\tiny{GB}}} \geq \lambda_{\text{\tiny{GB}}}^ {\text{\tiny{shear}}}
= - \frac{(D+1)(D-3)}{16}.
\label{boundshear}
\end{equation}
Note that for $D=5$ and 7, we recover the bounds
$\lgb^{\text{\tiny{shear}}}= -3/4$ and --2 which were originally found
in \cite{RobAlex,hofman} and \cite{Jan}, respectively. We also observe
that eq.~\reef{boundshear} precisely matches the bound \reef{cst2b}
that was derived to avoid the appearance of negative energy fluxes in
the vector (or spin 1) channel in section \ref{constraints}.

As $D$ becomes very large, we see that $\lgb^{\text{\tiny{shear}}}$ is
unbounded from below. This is a clear indication that
\eqref{boundshear} cannot be the correct lower bound for the full GB
theory.  Indeed, we now show that the correct lower bound comes from
the analysis in the sound channel.

\subsection{Sound channel}
The leading potential  (as $\hbar\to 0$) in this channel is given by the following expression
\begin{align}
U^0_ {\text{\tiny{sound}}}=&\left[(D-2)\left(1-2\lambda_{\text{\tiny{GB}}} f \right)^2\fin \left(D-3-2(D-1)\lambda_{\text{\tiny{GB}}} -2(D-5)\lambda_{\text{\tiny{GB}}} f \left(1-\lambda_{\text{\tiny{GB}}} f\right)\right)\right]^{-1}  \notag \\
&\times\Bigl[(D-2)(D-3)-6(D-1)(D-2)\lambda_{\text{\tiny{GB}}}+12(D-1)^2 \lambda_{\text{\tiny{GB}}}^2 + \notag \\
&\quad- 2\lambda_{\text{\tiny{GB}}} f \left(1-\lambda_{\text{\tiny{GB}}} f \right) \bigl((D-2)(D-9)+12(D-1)\lambda_{\text{\tiny{GB}}} + \notag \\
&\quad- 2(D-3)(D-5)\lambda_{\text{\tiny{GB}}} f \left(1-\lambda_{\text{\tiny{GB}}} f \right)\bigr) \Bigr] f\,.
\labell{U0sound}
\end{align}
The behaviour of this function is the same as that described in the
shear channel, as can be seen in figure \ref{figU0sound}.
\EPSFIGURE{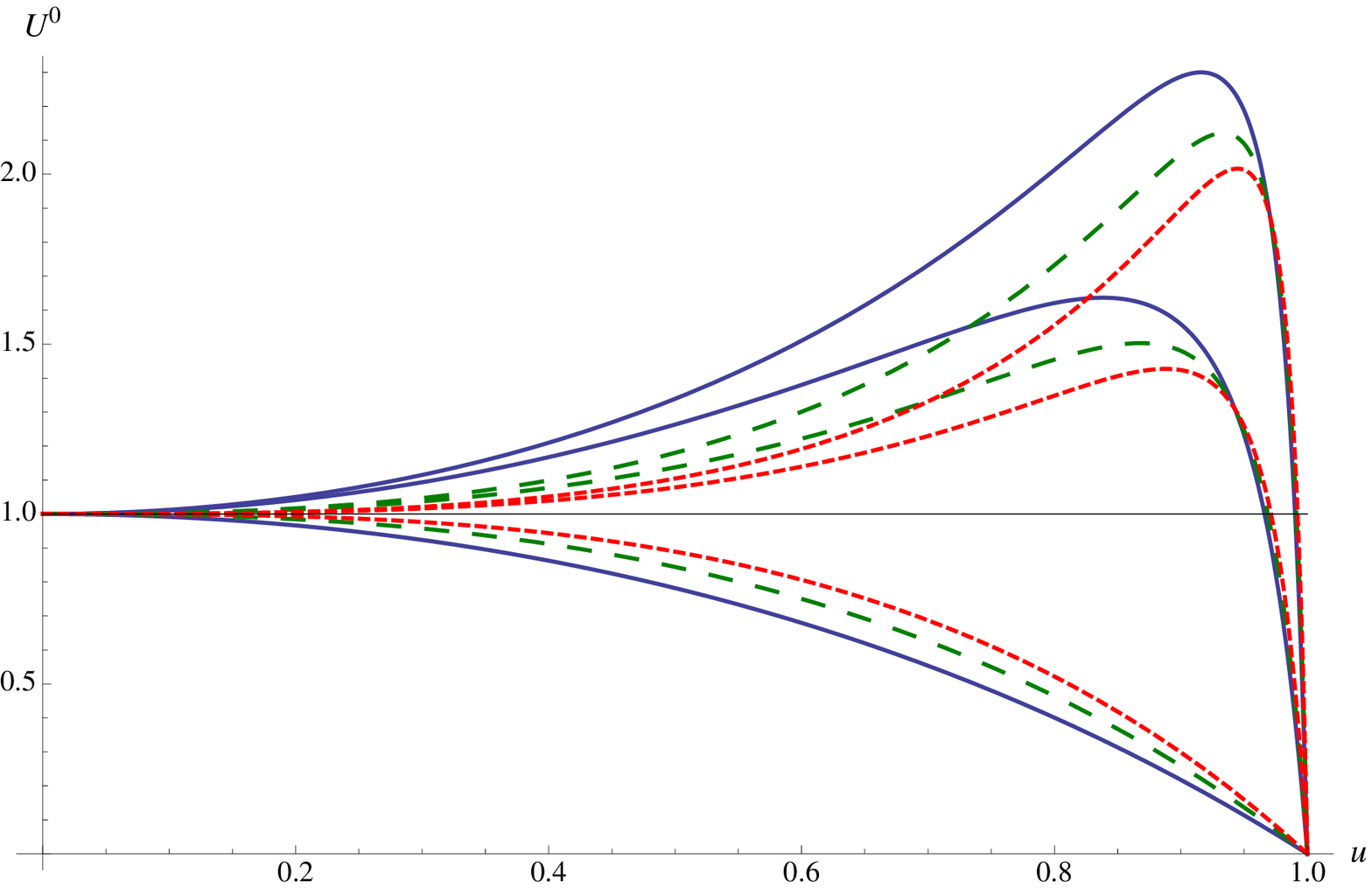,width=12cm} {(Colour online) Leading potential $U^0$ for the
Schr\"{o}dinger-like equation in the sound channel.  The blue (solid),
green (long dash) and red (small dash) lines correspond to $D=5$, $6$
and $7$, respectively.  For each dimension, the three potentials
correspond to $\lgb=-0.02,-1.5,-3.5$, from bottom to top. The behaviour
for higher $D$ is similar.\label{figU0sound}}

Following the same steps as in the previous case, we obtain the
following critical value for the GB coupling
\begin{equation*}
\lambda_{\text{\tiny{GB}}}^ {\text{\tiny{sound}}}= - \frac{(D-3)(3D-1)}{4(D+1)^2}.
\end{equation*}
Hence, by the same arguments used before to avoid causality violation
in the theory, we get the following lower bound
\begin{equation}
\lambda_{\text{\tiny{GB}}} \geq \lambda_{\text{\tiny{GB}}}^ {\text{\tiny{sound}}}
= - \frac{(D-3)(3D-1)}{4(D+1)^2}.
\label{boundsound}
\end{equation}
Again, for $D=5$ and 7, we recover $\lgb^{\text{\tiny{sound}}}= -7/36$
and --5/16, as was originally obtained in \cite{RobAlex,hofman} and
\cite{Jan}, respectively. Again comparing to the results in section
\ref{constraints}, we also find that eq.~\reef{boundsound} precisely
matches the bound \reef{cst3c} necessary to avoid the appearance of
negative energy fluxes in (what was denoted there as) the scalar or
spin 0 channel.

Eq.~\reef{boundsound} provides a more stringent lower bound than that
in the shear channel \reef{boundshear}. In particular, when $D$ becomes
very large, $\lgb^{\text{\tiny{sound}}}$ is bounded from below by
$-3/4$.

\subsection{Scalar channel}
For completeness, we present the analysis for the scalar channel, which
was already considered in \cite{Sin}.  The expression for the leading
potential is in this case
\begin{align}
U^0_ {\text{\tiny{scalar}}}=&\left[(D-4)\left(1-2\lambda_{\text{\tiny{GB}}} f\right)^2\fin \left(D-3-2(D-1)\lambda_{\text{\tiny{GB}}} - 2(D-5)\lambda_{\text{\tiny{GB}}} f (1-\lambda_{\text{\tiny{GB}}}f)\right)\right]^{-1}  \notag \\
&\times\Bigl[(D-3)(D-4)-2(D-1)(D-6)\lambda_{\text{\tiny{GB}}} - 4(D-1)^2 \lambda_{\text{\tiny{GB}}}^2 + \notag \\
&\quad- 2\lambda_{\text{\tiny{GB}}}f \left(1-\lambda_{\text{\tiny{GB}}} f \right) \bigl[3\left(D(D-7)+14\right)-4(D-1)(2D-7)\lambda_{\text{\tiny{GB}}} + \notag \\
&\quad-2(D-5)(D-7)\lambda_{\text{\tiny{GB}}}f \left(1-\lambda_{\text{\tiny{GB}}} f \right)\bigr]\Bigr] f.
\labell{U0scalar}
\end{align}
The behaviour of this function is shown in figure \ref{figU0scalar},
where we have made use of the bounds found in the shear and sound
channel to choose the minimum values that $\lambda_{\text{\tiny{GB}}}$
can take.
\EPSFIGURE{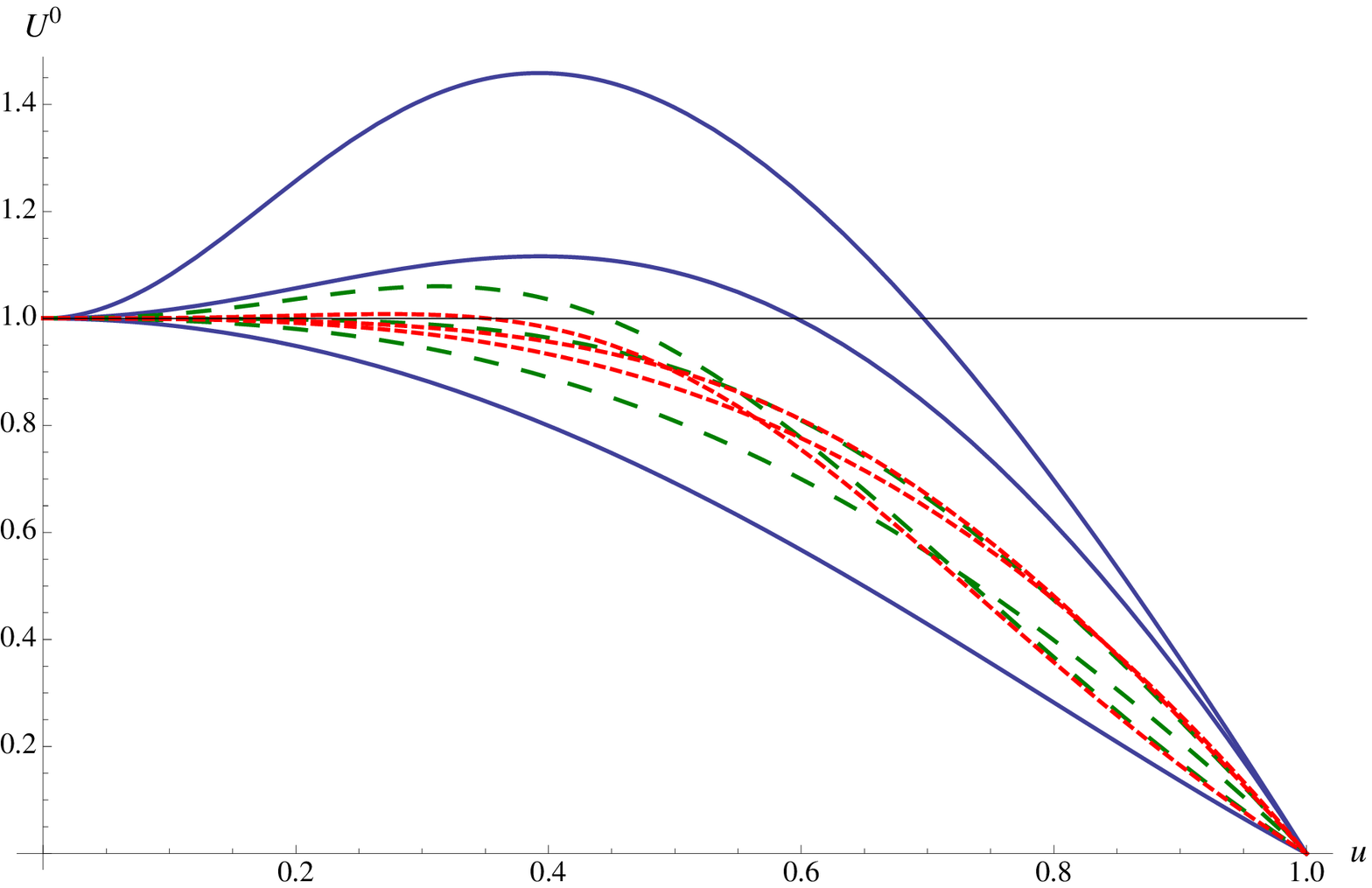,width=12cm} {(Colour online) Leading potential $U^0$ for the
Schr\"{o}dinger-like equation in the scalar channel.  The blue (solid),
green (long dash) and red (small dash) lines correspond to $D=5$, $6$
and $7$, respectively.  For value of $D$, the three potentials
correspond to $\lgb=-0.05, 0.15, 0.21$, from bottom to top.
\label{figU0scalar}}

We note that, as opposed to  the two previous cases, here the single
maximum in the potential appears when we increase the value of
$\lambda_{\text{\tiny{GB}}}$. The critical value of the coupling in
this channel is given by
\begin{equation*}
\lambda_{\text{\tiny{GB}}}^ {\text{\tiny{scalar}}}=  \frac{(D-3)(D-4)(D^2-3D+8)}{4(D^2-5D+10)^2}.
\end{equation*}
Hence, in this case this value gives an upper bound on the GB coupling
\begin{equation}
\lambda_{\text{\tiny{GB}}} \leq \lambda_{\text{\tiny{GB}}}^ {\text{\tiny{scalar}}}=
\frac{(D-3)(D-4)(D^2-3D+8)}{4(D^2-5D+10)^2}.
\label{boundscalar}
\end{equation}
For $D=5$ and 7, we recover the results $\lgb^{\text{\tiny{scalar}}} =
9/100$ and 3/16, first obtained in \cite{RobShenker2} and \cite{Jan},
respectively. Eq.~\reef{boundsound} also precisely matches the upper
bound \reef{cst1a} necessary to avoid the appearance of negative energy
fluxes in the tensor or (spin 2) channel. Further, we see that as $D$
becomes very large, $\lgb^{\text{\tiny{scalar}}}$ approaches the
critical value $\lgb=1/4$ from below.

\subsection{Plasma instabilities} \label{instab}
To complete this section, let us make some comments on other possible
instabilities in the theory.  So far, our analysis of the effective
 Schr\"{o}dinger equation \reef{schrodfinal} concerned
the possible appearance of superluminal signals in the dual CFT.
However, it was noted in \cite{RobShenker} that in $D=5$ GB gravity, a
new instability arises in the scalar channel at
$\lambda_{\text{\tiny{GB}}}=-1/8$. If one decreases the coupling below
this value, the potential $U^0_ {\text{\tiny{scalar}}}$ develops a
negative minimum located right in front of the horizon $u=1$. Following
\cite{RobRowan}, it was argued that for large $\textswab{q}$, this well
supports unstable quasinormal modes. This effect was extensively
studied for charged black holes in GB gravity in \cite{Sin,unstable}.
In particular, the results of \cite{Sin} indicate that a similar well
develops in the scalar channel of GB gravity for general $D$, \eg see
figure \ref{scalarinst}. To obtain the value at which this new
instability occurs, expand \eqref{U0scalar} to around $u=1$
\begin{align*}
U^0_ {\text{\tiny{scalar}}}=&-\frac{\left(1+\sqrt{1-4\lambda_{\text{\tiny{GB}}}} \right)\left((D-3)(D-4)-2(D-1)
(D-6)\lambda_{\text{\tiny{GB}}}-4(D-1)^2 \lambda_{\text{\tiny{GB}}}^2\right)}{2(D-4)\left(D-3-2(D-1)
\lambda_{\text{\tiny{GB}}}\right)}\notag\\
&\times\, \left(u^{\frac{D-1}2}-1\right)+\cdots\,.
\labell{outside}
\end{align*}
The sign of the coefficient of $\left(u^{(D-1)/2}-1\right)$ will
determine whether the potential develops a negative well in front of
the horizon. One finds that the negative well develops for \cite{Sin}
\begin{equation}
\lambda_{\text{\tiny{GB}}} < -\frac{(D-6)+\sqrt{5D(D-8)+84}}{4(D-1)}.
\labell{instscalar}
\end{equation}
Comparing this expression with the lower bound on $\lgb$ obtained in
the sound channel \eqref{boundsound}, we find that this instability
arises within the well-behaved regime only for $D=5$ and 6. That
is, for
\begin{align}
D&=5: \qquad -\frac{7}{36} \leq  \lambda_{\text{\tiny{GB}}} < -\frac{1}{8}, \notag \\
D&=6: \qquad -\frac{51}{196} \leq  \lambda_{\text{\tiny{GB}}} < -\sqrt{\frac{3}{50}}
\labell{unstable9}
\end{align}
the GB theory is dual to a well-behaved CFT, but the plasma still seems
to exhibit instabilities corresponding to unstable quasinormal modes in
the black hole background. Given that these modes have large momentum,
it seems that the instability causes the homogeneous CFT plasma to
``clump'' into an inhomogeneous configuration. However, for $D\ge7$ (or
$d\ge6$), the values given by \eqref{instscalar} are ruled out by the
lower bound \eqref{boundsound} and so no such scalar channel
instabilities arise in the dual CFT's (which are not already
pathological for other reasons).
\EPSFIGURE{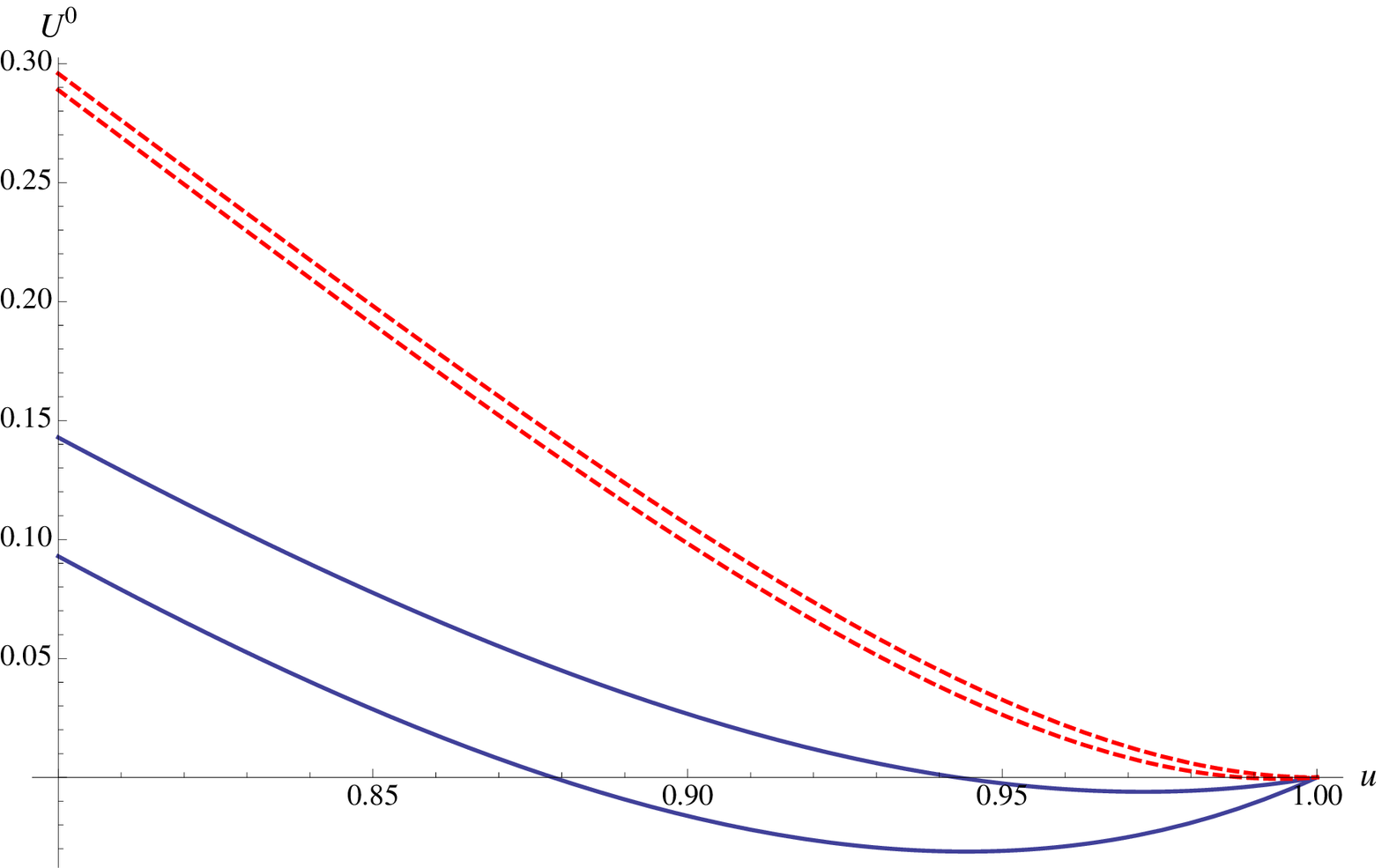,width=12cm} {(Colour online) Instabilities in $U^0_
{\text{\tiny{scalar}}}$.  The blue (solid) lines correspond to $D=5$
for  $\lambda_{\text{\tiny{GB}}}=-0.15,-0.19$, whereas the red (dashed)
lines correspond to $D=6$ for $\lambda_{\text{\tiny{GB}}}=-0.245,-0.26$
(the well is barely noticeable in this case). \label{scalarinst}}

Of course, one should also examine the shear and sound channels for
similar instabilities. In \cite{RobAlex}, it was observed for $D=5$
that the shear channel does not produce any new instabilities.
Similarly, examining the potential \eqref{U0shear}  the same result
applies for arbitrary values of $D$. It was also observed in
\cite{RobAlex} that in $D=5$ the sound channel can develop
instabilities for $\lambda_{\text{\tiny{GB}}} > 1/8$. Hence, this
effect only occurs outside of the physical regime, \ie $\lgb\le9/100$.
For $D\ge6$ we examine the potential \eqref{U0sound} for the possible
appearence of a negative well. We find $U^0_ {\text{\tiny{sound}}}\ge0$
everywhere in the range $1\ge u\ge0$ for $D\ge6$. Hence if we restrict
the GB coupling, as in eq.~\reef{range0}, to produce a physically
well-behaved CFT, the shear and sound channels do not produce any new
instabilities for a homogeneous plasma.

\section{Second-order hydrodynamics} \label{gohydro}

Hydrodynamics of relativistic fluids is discussed in
\cite{SonRomat,Romat2}.  Here we summarize a few relevant results for
our computations.  Let us consider a $(D-1)$-dimensional conformal
fluid, as is appropriate for the CFT dual of $D$-dimensional GB
gravity. The dynamics of the long wavelength fluctuations can be
organized in terms of a derivative expansion. In particular, in the
absence of any conserved charges, the dynamics of the hydrodynamic
modes is simply governed by the conservation of the stress-energy tensor:
$\nabla_a T^{ab}=0$. The latter includes both the equilibrium part,
involving the local energy density $\varepsilon$ and pressure $P$, and
a dissipative part $\Pi^{ab}$
\begin{equation}
T^{ab}=\varepsilon \, u^a u^b+P\Delta^{ab}+\Pi^{ab}\,,\labell{lead3}
\end{equation}
where $\Delta^{ab}=g^{ab}+u^a u^b$ and $u^a$ is the local four-velocity
of the fluid, with $u^a u_a=-1$. For a conformal fluid, we also have
the restriction that the trace of the stress tensor must vanish, \ie
$T^a{}_a=0$. In the equilibrium contribution above, this requires that
$P=\varepsilon/(D-2)$.

The dissipative term can be written as an infinite series expansion in
velocity gradients and curvatures (for a fluid in a curved background),
where the coefficients in this expansion are the transport coefficients
of the fluid. At first order for a conformal fluid, one has
\begin{equation*}
\Pi^{ab}_1 =-\eta \, \sigma^{ab}\,,
\end{equation*}
where
\begin{equation*}
\sigma^{ab}=2\nabla^{\langle a}u^{b \rangle} \equiv \Delta^{ac} \Delta^{bd}
\left(\nabla_c u_d + \nabla_d u_c \right)-\frac{2}{D-2}\Delta^{ab} \Delta^{cd}
\nabla_c u_d\,.
\end{equation*}
Implicitly here, we have used the condition that $T^a{}_a=0$ to
eliminate any bulk viscosity contribution. Hence the only nonvanishing
transport coefficient appearing at this order is the shear viscosity
$\eta$. Of course, if we truncate the hydrodynamic equations to include
only this ``Newtonian'' first-order term, there are superluminal modes
propagating in the fluid \cite{Hiscock}.  One can try to avoid this
problem by going to next order in the expansion \cite{MIS}. In general
for a conformal fluid, five new transport coefficients will appear in
the second-order term $\Pi^{ab}_2$. However, there is only one
contribution which is independent of both the vorticity and background
curvature, as well as linear,
\begin{equation*}
\Pi^{ab}_2 =\eta \tau_\Pi\left[^{\langle} u \cdot \nabla \sigma^{ab \rangle}+\frac{1}{D-2}
 \sigma^{ab} \left(\nabla \cdot u\right)\right] \,.
\end{equation*}
The additional transport coefficient above is the relaxation time
$\tau_\Pi$ and this linear term is sufficient to examine the question
of causality within the hydrodynamic regime \cite{Muronga}.

In \cite{SonRomat}, the authors studied linearized fluctuations of the
second-order hydrodynamics of conformal fluids and found the following
results:

\subsubsection*{$\bullet$ Shear channel}
The dispersion relation in the shear channel was found to be
\begin{equation}
-\textswab{w}^2 \tau_\Pi T-\frac{i \, \textswab{w}}{2\pi}
+ \textswab{q}^2 \frac{\eta}{s}=0,
\labell{shearhydro1}
\end{equation}
where $\textswab{w}$ and $\textswab{q}$ are defined in eq.~\reef{qw}.
We are interested in the hydrodynamic limit in which $\textswab{w},
\textswab{q} \to 0$, with $\textswab{w}/\textswab{q}$ kept fixed. In
this limit, the dispersion relation \reef{shearhydro1} yields the
following Taylor series solution
\begin{equation}
\textswab{w}=-2\pi i \, \frac{\eta}{s} \, \textswab{q}^2 -8 \pi^3 i \, \tau_\Pi T
 \, \frac{\eta^2}{s^2} \, \textswab{q}^4 + \mathcal{O}(\textswab{q}^6).
\labell{shearhydro2}
\end{equation}
Note that we have discarded a solution where $\textswab{w}$ remains
finite as $\textswab{q} \to 0$, hence lying beyond the hydrodynamic
regime  --- see \cite{SonRomat}. The wave-front speed with which
disturbances propagate out from a discontinuity in any initial data is
governed by \cite{Fox}
\begin{equation}
v_{\text{\tiny{shear}}}^{\text{\tiny{front}}} \equiv \lim_{\vert \textswab{q} \vert
\to \infty} \frac{\mathop{\rm Re} (\textswab{w})}{\textswab{q}}
=\sqrt{\frac{\eta}{\tau_\Pi T \, s}}\,. \labell{vfrontshear}
\end{equation}
Thus, requiring causality to be preserved in this channel of a
conformal fluid imposes the following bound \cite{RobAlex}
\begin{equation}
\tau_\Pi T \geq \frac{\eta}{s}\,.
\labell{bound1}
\end{equation}

\subsubsection*{$\bullet$ Sound channel}
The dispersion relation in the sound channel was found to be
\begin{equation}
-\textswab{w}^3 \tau_\Pi T-\frac{i \, \textswab{w}^2}{2\pi} +
\textswab{w} \, \textswab{q}^2 c_s^2 \tau_\Pi T + \textswab{w} \,
\textswab{q}^2 \frac{2(D-3)}{(D-2)}\frac{\eta}{s}+\frac{i \, \textswab{q}^2 c_s^2}{2\pi}=0,
\labell{soundhydro1}
\end{equation}
where $c_s$ is the speed of sound, which in any conformal fluid is a
constant
\begin{equation}
c_s^2=\frac{1}{D-2}\,.
\label{cs1}
\end{equation}
At small momentum $\textswab{q}$, the Taylor series solution to the
above equation corresponding to the sound wave is
\begin{equation}
\textswab{w}=c_s \, \textswab{q}-2\pi i \, \Gamma T \, \textswab{q}^2
+ \frac{4 \pi^2}{c_s} \Gamma T \left(c_s^2 \tau_\Pi T- \frac{\Gamma T}{2}\right)
\textswab{q}^3 + \mathcal{O}(\textswab{q}^4),
\labell{soundhydro2}
\end{equation}
where
\begin{equation}
\Gamma T = \frac{D-3}{D-2} \, \frac{\eta}{s}\,.
\labell{gammaTT}
\end{equation}
Note that we have only written one of the three solutions to
\eqref{soundhydro1}.  The second solution corresponds to waves
propagating in the opposite direction, \ie replace $\textswab{q}$ with
$-\textswab{q}$ in eq.~\reef{soundhydro2}. The third solution again lies
beyond the regime of validity of hydrodynamics. Given
eq.~\reef{soundhydro2}, we have
\begin{equation}
v_{\text{\tiny{sound}}}^{\text{\tiny{front}}} \equiv \lim_{\vert \textswab{q}
\vert \to \infty} \frac{\mathop{\rm Re} (\textswab{w})}{\textswab{q}}
=\sqrt{c_s^2+\frac{2(D-3)}{(D-2)}\frac{\eta}{s}\frac{1}{\tau_\Pi T}}\,.
\labell{vsound0}
\end{equation}
Hence, causality in this channel imposes the following bound
\begin{equation}
\tau_\Pi T \geq 2 \frac{\eta}{s}.
\label{bound2}
\end{equation}
It is quite interesting that despite the explicit appearance of $D$ in
eq.~\reef{vsound0}, the bound \reef{bound2} is independent of the
dimension and takes the same simple form as originally found for $D=5$
in \cite{RobAlex}.

We see that in both channels considered above, the wave-front speed of
the linearized fluctuations diverges as the relaxation time $\tau_\Pi$
goes to zero.  Notice also that the front velocity in the sound channel
is always larger than that in the shear channel.  Therefore, the bound
\eqref{bound2} provides a stronger constraint on the transport
coefficients.  As emphasized in \cite{RobAlex}, the constraints
\eqref{bound1} and \eqref{bound2} arise when considering linearized
modes outside the regime of validity of hydrodynamics.  Hence, these
constraints must not be regarded as fundamental, as they merely
indicate where a certain approximate framework describing the conformal
fluid becomes problematic.

\subsection{Holographic Gauss-Bonnet hydrodynamics}

Next we would like to apply the now standard techniques
\cite{ssmin,SonRomat} of the AdS/CFT framework to calculate the
transport coefficients $\eta$ and $\tau_\Pi$ for the CFT's dual to GB
gravity in an arbitrary spacetime dimension. The shear viscosity for GB
gravity in $D=5$ was originally calculated in \cite{RobShenker},
however, these calculations are easily extended to GB gravity in any
dimension \cite{RobShenker,Sin}. Their result is
\begin{equation}
\frac{\eta}{s}= \frac{1}{4\pi} \left[1-\frac{2(D-1)}{(D-3)}
\lambda_{\text{\tiny{GB}}} \right]\,,
\labell{etas0}
\end{equation}
where the entropy density $s$ is given in eq.~\reef{entropy}. We
reproduce this result with our analysis of the sound waves below.

Hence it remains for us to calculate the relaxation time $\tau_\Pi$.
For this purpose, we must solve the equation of motion
\eqref{soundeom}, which describes the propagation of sound waves in the
CFT plasma dual to GB gravity.  This allows us to obtain expressions
that can be compared to those presented in the previous subsection.
First, we make the following ansatz for the solution near the horizon
\begin{equation*}
Z_{\text{\tiny{sound}}}= (1-u^{\frac{D-1}{2}})^\beta,
\end{equation*}
where there could be an overall coefficient independent of $u$, which
is not relevant for our discussion. If we replace this ansatz into
\eqref{soundeom}, we readily obtain the following two solutions
\begin{equation*}
\beta= \pm \frac{i\textswab{w}}{2},
\end{equation*}
with the negative sign satisfying the correct infalling boundary
condition at the horizon.  Let us now focus on the hydrodynamic limit,
where $\textswab{w}$, $\textswab{q} \rightarrow 0$ with
$\frac{\textswab{w}}{\textswab{q}}$ kept fixed.  We will write our
solution as a series in $\textswab{q}$ and then solve
eq.~\eqref{soundeom} perturbatively.  The horizon boundary condition
implies that we can write our solution as
\begin{equation}
Z_{\text{\tiny{sound}}}= (1-u^{\frac{D-1}{2}})^{-\frac{i \textswab{w}}{2}}
 \left(z_0(u; \textswab{w}, \textswab{q}) + i \textswab{q} \, z_1(u; \textswab{w},
 \textswab{q}) + \textswab{q}^2 z_2(u; \textswab{w}, \textswab{q}) + \mathcal{O}(\textswab{q}^3) \right),
\end{equation}
where at the horizon $\lim_{u \to 1} z_i (u; \textswab{w},
\textswab{q}) = \delta^0_i$ with $i=0,1,2$. Further each of the $z_i
(u; \textswab{w}, \textswab{q})$ should only depend on the ratio
$\textswab{w}/\textswab{q}$ which implies a scaling invariance
 \beq
z_i (u; \mu \, \textswab{w}, \mu \, \textswab{q})= z_i (u;
\textswab{w}, \textswab{q}) \labell{scale78}
 \eeq
for any $\mu$.  In this expansion, we have only written terms to the
order necessary to identify the sound wave dispersion relation
\eqref{soundhydro2}, which is obtained by imposing the Dirichlet
boundary condition on $Z_{\text{\tiny{sound}}}$ at the asymptotic
boundary
\begin{equation}
\lim_{u \to 0} Z_{\text{\tiny{sound}}}=0\,.
\labell{dirichlet}
\end{equation}
By comparing the terms of the dispersion relation at each order in
$\textswab{q}$, we will be able to identify the expressions for the
conformal sound speed $c_s$ (the leading order), the shear viscosity to
entropy ratio $\eta/s$ (the first order) and the relaxation time $\tau_\Pi$
(the second order). The analysis is greatly simplified by adopting a new
radial coordinate
\begin{equation}
x \equiv \sqrt{1-4\lambda_{\text{\tiny{GB}}}(1-u^{\frac{D-1}{2}})}.
\end{equation}
With this radial coordinate, the horizon and asymptotic boundary
conditions become
\begin{align}
\lim_{x \to 1} z_i (x; \textswab{w}, \textswab{q}) &= \delta^0_i \qquad \text{(horizon)}, \notag \\
\lim_{x \to (1-4\lambda_{\text{\tiny{GB}}})^{1/2}} Z_{\text{\tiny{sound}}} &= 0 \qquad \text{(boundary)}.
\label{bcsx}
\end{align}

\subsubsection{Speed of sound}

Solving eq.~\eqref{soundeom} perturbatively, the leading order term
gives a second-order differential equation for $z_0(x)$.  We will not
write the explicit equation here, but merely state its solution,
\begin{equation}
z_0= \frac{ \textswab{q}^2 \left[(D-5)x^2 + 4x -(D-1) \left(1-4
\lambda_{\text{\tiny{GB}}}\right) \right]-8\left(D-2\right)\fin
\textswab{w}^2 \lambda_{\text{\tiny{GB}}} \, x}{4 \left[(D-1)
\textswab{q}^2 -2\left(D-2\right)\fin\textswab{w}^2\right]
\lambda_{\text{\tiny{GB}}} x}\,.
\labell{solz0}
\end{equation}
By imposing the Dirichlet boundary condition \eqref{bcsx} and solving
for \textswab{w}, we find the expected dispersion relation
\begin{equation}
\textswab{w}= \frac{1}{\sqrt{D-2}}\, \textswab{q}\,.
\end{equation}
Of course, the factor $1/\sqrt{D-2}$ is precisely the speed of sound
\reef{cs1} for a conformal fluid in $D-1$  dimensions.

\subsubsection{Shear viscosity-to-entropy ratio}

To first order in $\textswab{q}$ we obtain a second-order differential
equation for $z_1(x)$, which involves the previously found $z_0(x)$.
This equation is even more involved than that for $z_0(x)$. To proceed we
make the following ansatz for the solution
$z_1=\left(\textswab{w}/\textswab{q}\right)\, z_0(x) \, F(x)$.  Again,
it is not very illuminating to write the resulting equation and so we
simply give its solution
\begin{align}
F(x)=&\frac{1} { \textswab{q}^2 \left[(D-5)x^2 + 4x -(D-1) \left(1-4\lambda_{\text{\tiny{GB}}}\right) \right]-8\left(D-2\right)\fin\textswab{w}^2 \lambda_{\text{\tiny{GB}}} \, x} \times \notag \\
&\Biggl[ \left[ \textswab{q}^2 \left[(D-5)x^2 + 4x -(D-1) \left(1-4\lambda_{\text{\tiny{GB}}}\right) \right]-8\left(D-2\right)\fin\textswab{w}^2 \lambda_{\text{\tiny{GB}}} \, x\right] \times \notag\\
&\biggl[\sqrt{\frac{(D-1)(1-4\lambda_{\text{\tiny{GB}}})}{4(D-5)}} \arctan \left(\frac{\sqrt{(D-1)(D-5)(1-4\lambda_{\text{\tiny{GB}}})}} {(D-1)(1-4\lambda_{\text{\tiny{GB}}})+(D-5)x}(1-x)\right) + \notag \\
&+\frac{1}{2} \ln \left(\frac{1+x}{2}\right) - \frac{(D-1)(1-4\lambda_{\text{\tiny{GB}}})}{4(D-5)} \ln \left(\frac{(D-5)x^2+(D-1) (1-4\lambda_{\text{\tiny{GB}}})}{2(D-3)-4(D-1)\lambda_{\text{\tiny{GB}}}}\right) \biggr] + \notag \\
&-4\textswab{q}^2 x(1-x)\left((D-3)-2(D-1)\lambda_{\text{\tiny{GB}}}\right) \Biggr]\,
.\labell{bigF}
\end{align}
Note that while at first sight it might seem that this result is not
well-behaved at $D=5$,  taking the limit $D \to 5$ carefully, recovers
the results obtained in \cite{RobAlex} --- in taking this limit, one
should recall that $\lambda_{\text{\tiny{GB}}} <1/4$. Given this
general solution \reef{bigF}, we can impose the Dirichlet boundary
condition at infinity and solve for $\textswab{w}$, obtaining the
following dispersion relation up to second-order in \textswab{q}
\begin{equation*}
\textswab{w}=\frac{1}{\sqrt{D-2}} \textswab{q} -i \frac{(D-3)}{2(D-2)}
 \left[1-\frac{2(D-1)}{(D-3)}\lambda_{\text{\tiny{GB}}}\right]\textswab{q}^2.
\end{equation*}
By comparing to eqs.~\eqref{soundhydro2} and \reef{gammaTT}, we see
that the factor in brackets in the second-order coefficient is
essentially the ratio of the shear viscosity to entropy density with
\begin{equation}
\frac{\eta}{s}= \frac{1}{4\pi} \left[1-\frac{2(D-1)}{(D-3)}\lambda_{\text{\tiny{GB}}}\right].
\label{etas}
\end{equation}
Of course, this precisely reproduces the result found previously in
\cite{RobShenker,Sin}.

\subsubsection{Relaxation time and causality violation}

\begin{figure}
\centering
\epsfig{file=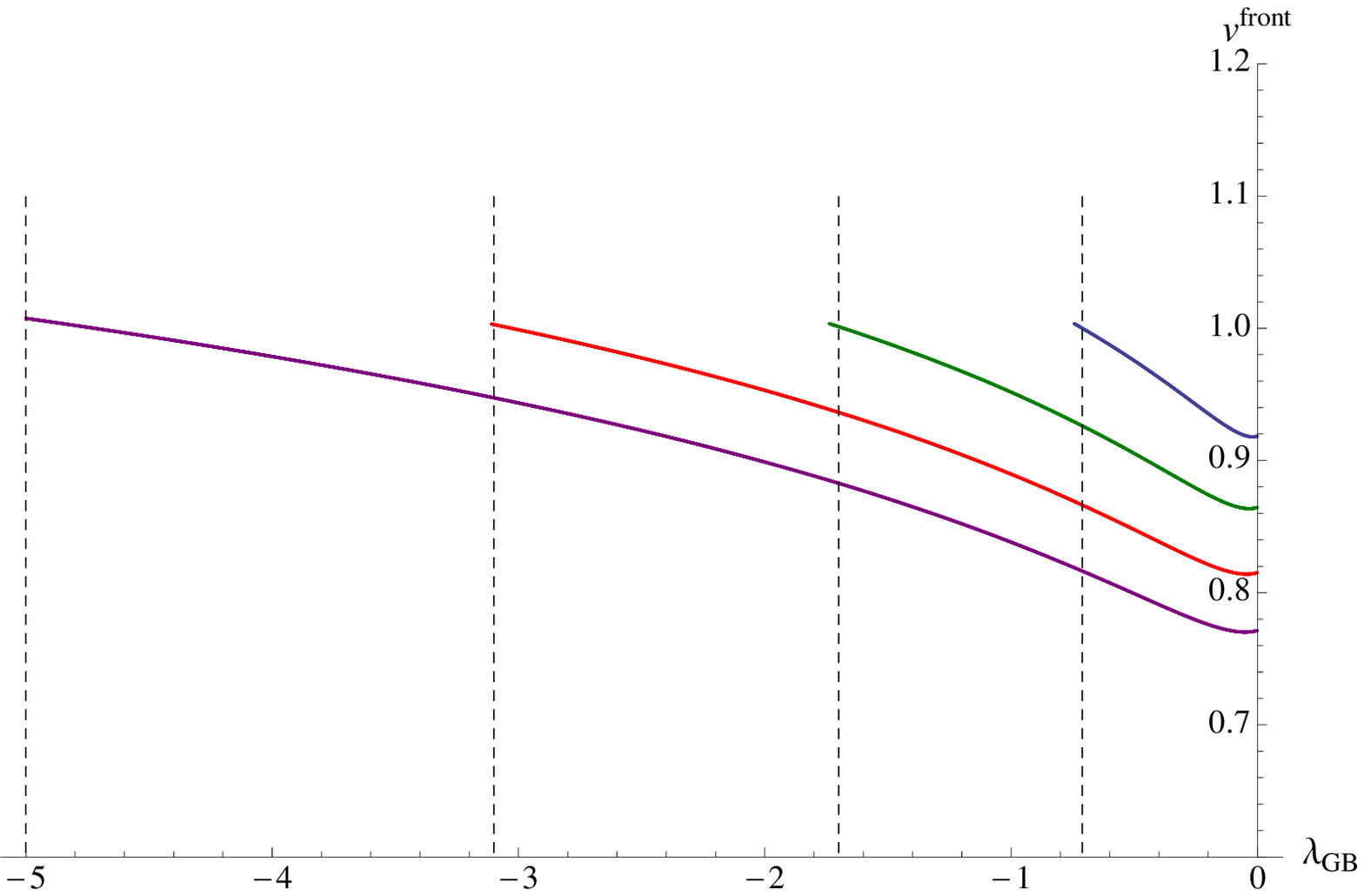,width=7.5cm}
\epsfig{file=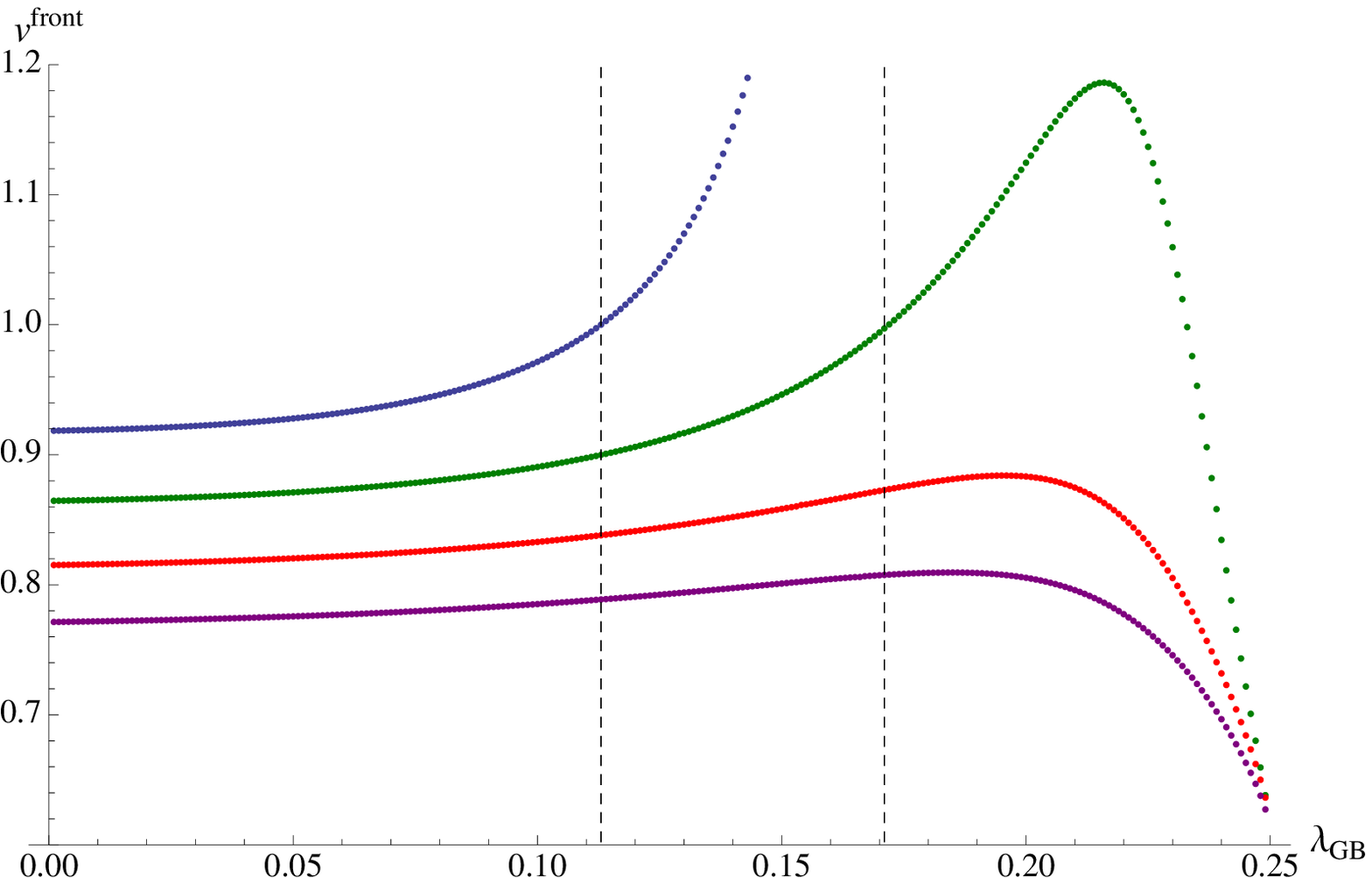,width=7.5cm}
\caption{(Colour online) Front velocity in the sound channel.  From top to bottom,
the blue, green, red and purple curves correspond to $D=5$, $6$, $7$ and $8$, respectively.
The dashed vertical lines indicate where this velocity reaches one on
the various curves. \label{figvfront}}
\end{figure}

To second order in $\textswab{q}$ we obtain a differential equation for
$z_2(x)$, involving $z_0(x)$ and $z_1(x)$.  Unfortunately, we
were only able to solve this equation  numerically.\footnote{The
details of numerical calculations will not
be presented here, but are available from the authors upon request.}
Naturally, this means that we are not  able to obtain general
expressions for arbitrary values of $D$. Here we present numerical
results for $D=5,6,7,8$ but these are indicative of the behaviour for
higher dimensions as well. Note that the constraint \eqref{bound2} to
avoid causality violations requires
\begin{equation*}
\tau_\Pi T - 2 \frac{\eta}{s} \geq 0.
\end{equation*}
We show the results of our numerical computations for this quantity in
figure \ref{fighydro}.  We see that for $D>6$, causality does not
impose an upper bound on the GB coupling, however, we still require
$\lgb<1/4$ as was assumed from the very beginning of the discussion.
From these numerical results, we find the following bounds for having
causal second-order hydrodynamics,
\begin{align}
D&=5: \qquad -0.711 \leq  \lambda_{\text{\tiny{GB}}} \leq 0.113, \notag \\
D&=6: \qquad -1.7 \leq  \lambda_{\text{\tiny{GB}}} \leq 0.171, \notag \\
D&=7: \qquad -3.1 \leq  \lambda_{\text{\tiny{GB}}} < 0.25, \notag \\
D&=8: \qquad -5.0 \leq  \lambda_{\text{\tiny{GB}}} < 0.25.
\labell{boundshydro}
\end{align}
Note that we do not quote the bounds in higher dimensions with the same
accuracy because the numerical analysis became more unstable for large
values of $D$. In figure \ref{figvfront}, we also present the behaviour
of the front velocity \reef{vsound0} in the sound channel for these
various cases.
\EPSFIGURE{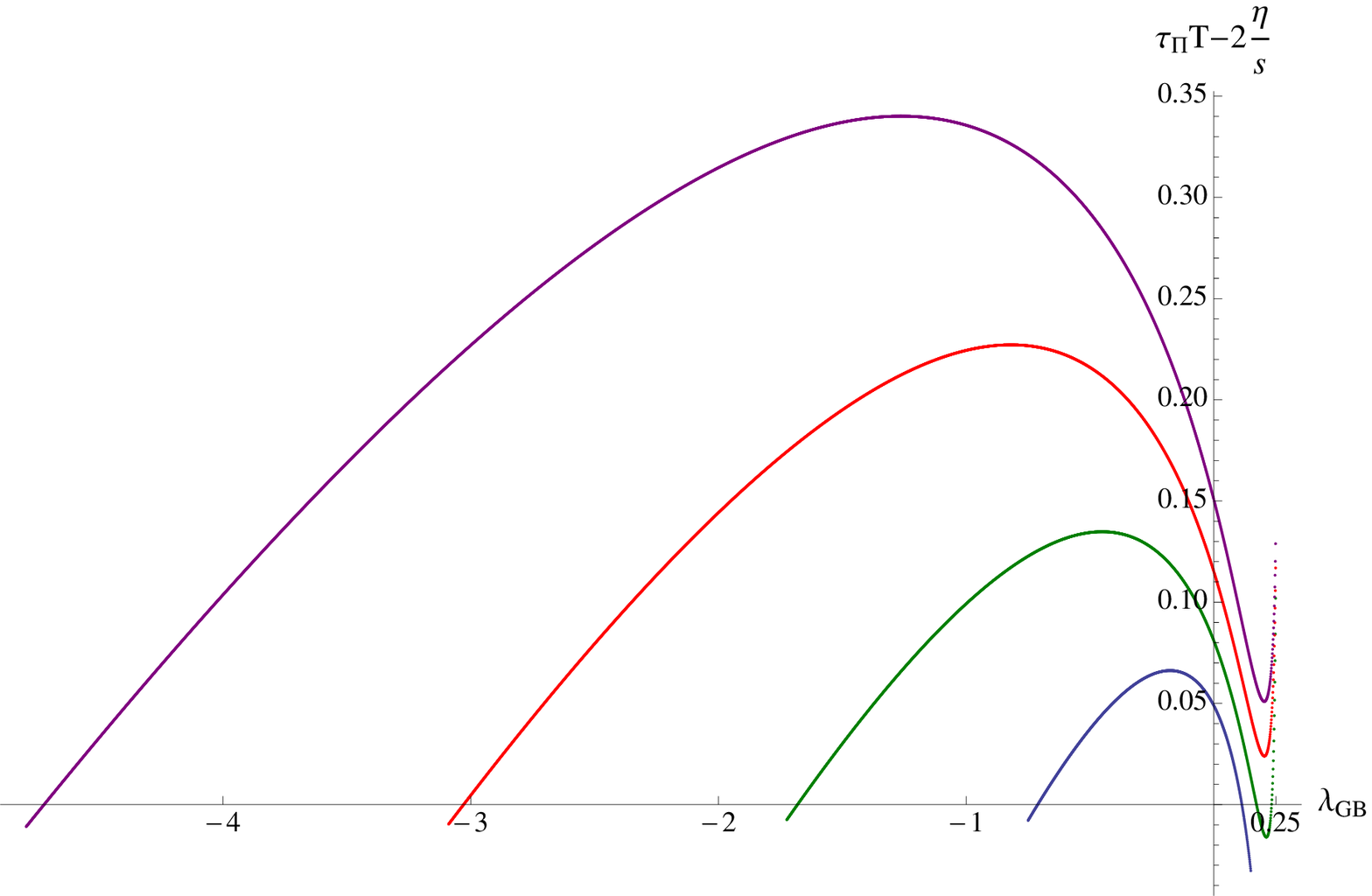,width=12cm} {(Colour online) Causality in the second-order GB
hydrodynamics is violated once the value of $\tau_\Pi T - 2
\frac{\eta}{s}$ becomes negative.  From bottom to top, the blue, green,
red and purple curves correspond to $D=5$, $6$, $7$ and $8$,
respectively. \label{fighydro}}
%

\section{Discussion} \label{discuss}

In this present paper, we have made a broad study of holographic GB
gravity in arbitrary dimensions. The first step was to construct the
AdS/CFT dictionary which applies for arbitrary $D$ in section
\ref{dictionary}. There are two dimensionless parameters which
characterize GB gravity, \ie the ratio of the AdS curvature scale to
Planck scale, $\tL/\lp$, and the GB coupling, $\lgb$. Since we are only
dealing with the gravitational sector of the AdS theory, these
parameters are naturally related to CFT parameters which determine the
behaviour of the $n$-point functions of the stress-energy tensor. So
for example, using eqs.~\reef{ctfinal} and \reef{t2holo}, we can
translate the gravitational parameters to expressions in terms of the
central charge $\ct$ and the scattering parameter $t_2$. Alternatively,
they could be expressed in terms of the three-point couplings
$\A,\B,\C$ using eqs.~\reef{central} and \reef{misha2} (keeping in mind
the constraint \reef{misha3}). Given the results of section
\ref{dictionary}, one can express any physical quantities, \eg
$\eta/s$, entirely in terms of parameters appearing in the CFT. This
is, of course, an essential step in the program of using holographic
models to characterize the properties of the quark-gluon plasma
\cite{holoqgp}.

Using these results, an interesting observation can be made with
regards to the relation between the central charge $\ct$ appearing in
the two-point function and the ``central charge'' characterizing the
entropy density. That is, one can use the entropy density of a CFT in
$d$ dimensions to provide an alternative definition of a central
charge: $C_S\propto s/T^{d-1}$. In two dimensions, $\ct$ and $C_S$ are
related by a simple numerical factor. However, there is no evidence of
any simple relation for strongly coupled CFT's in higher dimensions,
\eg see \cite{subir,fradkin}. This alternative central charge was
recently considered for CFT's dual to Einstein gravity in \cite{pavel}
and here we extend the discussion to consider GB gravity in an
arbitrary number of dimensions. First we adopt the normalization of
\cite{pavel} in defining $C_S$ as
 \beq
C_S\equiv \frac{d+1}{d-1}\left(\frac{d}{2\pi^{3/2}}\right)^d
\frac{\Gamma\left((d+1)/2\right)}{\sqrt{\pi}}\ \frac{s}{T^{d-1}}\,.
 \labell{centrals0}
 \eeq
Then examining the entropy density \reef{entropy} for GB gravity along
with the holographic expression for $\ct$ in eq.~\reef{ctfinal}, we see
that these two central charges have a rather complicated relation:
 \beq
C_S= \ct\
\frac{\left(1+2\gamma_d\,t_2\right)^d}{\left(1+\gamma_d\,t_2\right)^{d-1}}
\labell{centrals}
 \eeq
where $\gamma_d=(d-2)(d-3)/(4d(d-1))$. The normalization was chosen
above so that with $t_2=0$ (\ie for the CFT dual of Einstein gravity)
we have $C_S=\ct$. In general, one sees that $C_S$ is proportional to
$\ct$, however, it also has a highly nontrivial dependence on $t_2$.
Hence we are lead to conclude that these two central charges are
independent parameters in the family of CFT's dual to GB gravity.

We may go further by recalling that $t_2$ is constrained by
eqs.~(\ref{cstrn1}--\ref{cstrn3}). With $t_4=0$, combining the tensor
\reef{cstrn1} and scalar \reef{cstrn3} constraints yields
 \beq
-\frac{d-1}{d-3}\le t_2\le d-1\,.
 \labell{ranget2}
 \eeq
We may note that in this range \reef{ranget2}, the ratio $C_S/\ct$ is a
monotonically increasing function of $t_2$ for any dimension. Hence
these constraints in turn lead to
 \beq
\frac{3d+2}{4d}\,\left(\frac{2(d+2)}{3d+2}\right)^d \le \frac{C_S}{\ct}
\le \frac{d^2-d+6}{4d}\left(\frac{2(d^2-3d+6)}{d^2-d+6}\right)^d\ .
 \labell{rangeR}
 \eeq
Explicitly evaluating these results \reef{rangeR} for various
dimensions, we find
\begin{align}
d&=4: \qquad 0.4723 \leq  C_S/\ct \leq \ 1.7147\,, \notag \\
d&=5: \qquad 0.3220 \leq  C_S/\ct \leq \ 3.6714\,, \notag \\
d&=6: \qquad 0.2185 \leq  C_S/\ct \leq \ 8.4280\,, \notag \\
d&=7: \qquad 0.1477 \leq  C_S/\ct \leq 19.6317\,. \labell{rangeR1}
\end{align}
Here we see that the maximum possible value of $C_S/\ct$ grows
monotonically as $d$ increases while the minimum ratio is monotonically
decreasing and asymptotically approaches zero.

Following \cite{pavel}, we can compare these results for the strongly
coupled holographic field theories to those for free fields. First,
substituting the free field results \reef{freefield} for $\A$, $\B$ and
$\C$ into eq.~\reef{central} yields
 \beq
\ct=\frac{1}{\Omega_{d-1}^2}\left[\frac{d}{d-1}\,n_s+\frac{d}{2}\,\tilde{n}_f
+\frac{d^2}{2}\,\tilde{n}_t\right]\,,
 \labell{ctfree}
 \eeq
where $n_s$, $\tilde{n}_f$ and $\tilde{n}_t$ denote the number of
(massless) degrees of freedom contributed by scalar, fermion and tensor
fields, respectively, as described after eq.~\reef{freefield}. It
is straightforward to calculate the entropy density for free massless
fields, \eg see \cite{fradkin}. Then with the definition
\reef{centrals0}, we find
 \beq
C_S=2d(d+1)\left(\frac{d}{4\pi^2}\right)^d\Gamma(d-1)\,\zeta(d)\left[n_s
+\left(1-2^{1-d}\right)\,\tilde{n}_f+\tilde{n}_t\right]\,.
 \labell{csfree}
 \eeq
Combining these results, we arrive at
 \beqa
\left.\frac{C_S}{\ct}\right|_{free}&=&\frac{2}{\sqrt{\pi}}(d^2-1)
\left(\frac{d}{2\pi}\right)^d
\frac{\Gamma\left((d-1)/2\right)}{\Gamma(d/2)}\,\zeta(d)\, \frac{
n_s+\left(1-2^{1-d}\right)\,\tilde{n}_f+\tilde{n}_t}{n_s+\frac{d-1}{2}\,\tilde{n}_f
+\frac{d(d-1)}{2}\,\tilde{n}_t}
 \nonumber\\
&=&K(d)\, \left[1+\frac{d-3}{2(d-1)}\,t_2+\left(
\frac{1}{4\left(1-2^{-d}\right)} -\frac{2}{d^2-1}
\right)\,t_4\right]\,,
 \labell{freedR}
 \eeqa
where
 \beq
K(d)= \frac{16}{\sqrt{\pi}}\left(1-2^{-d}\right)
\left(\frac{d}{2\pi}\right)^d
\frac{\Gamma\left((d-1)/2\right)}{\Gamma(d/2)}\,\zeta(d)\,.
 \labell{Konstant}
 \eeq
For example, with $d=4$, this free field result becomes \cite{pavel}
 \beq
\left.\frac{C_S}{\ct}\right|_{free,d=4}=\frac{8}{3}\frac{
n_s+\frac{7}{8}\,\tilde{n}_f+2\,n_v}{n_s+\frac{3}{2}\,\tilde{n}_f
+12\,n_v}=\frac{4}{3} \left(1+{1\over6}t_2+\frac{2}{15}t_4\right)\,.
 \labell{free4R}
 \eeq
Note that in this case, the pre-factor in eq.~\reef{freedR} yields
$1/K(d=4)=3/4$, which corresponds to the ratio of the entropy density
of $N=4$ super-Yang-Mills at strong and at weak coupling \cite{amanda}.

If we focus on $t_4=0$, as for the dual of GB gravity, eq.~\reef{freedR}
reduces to the rather simple expression
 \be
\left.\frac{C_S}{\ct}\right|_{free}=K(d)\,
\left[1+\frac{d-3}{2(d-1)}\,t_2\right]\,,
 \labell{freedR2}
 \ee
which we might compare to the analogous result \reef{centrals} at
strong coupling. The constraints (\ref{cstrn1}--\ref{cstrn3}), derived
from considerations of the energy flux, make no reference to the
strength of the coupling and so eq.~\reef{ranget2} applies equally well
for the free field theories. Hence this constraint on $t_2$ in turn
leads to
 \beq
\frac{1}{2}\,K(d) \le \left.\frac{C_S}{\ct}\right|_{free} \le
\frac{d-1}{2}\,K(d)\ .
 \labell{rangeR3}
 \eeq
Explicitly evaluating these results \reef{rangeR3} for various
dimensions, we find
\begin{align}
d&=4: \qquad 0.6667 \leq  \left.C_S/\ct\right|_{free}\ \leq \ 2\,, \notag \\
d&=5: \qquad 1.0884 \leq  \left.C_S/\ct\right|_{free}\ \leq \ 4.3536\,, \notag \\
d&=6: \qquad 2.2781 \leq  \left.C_S/\ct\right|_{free}\ \leq \ 11.3906\,, \notag \\
d&=7: \qquad 5.7890 \leq  \left.C_S/\ct\right|_{free}\ \leq\ 34.7338\,, \labell{rangeR4}
\end{align}
which are readily compared with the corresponding results at strong
coupling in eq.~\reef{rangeR1}. Here we see that both the maximum and
minimum possible value of $\left.C_S/\ct\right|_{free}$ are growing
monotonically as $d$ increases. Further the ranges of $C_S/\ct$ for the
free fields and the strongly coupled holographic CFT partially overlap,
however, generally for a given value of $t_2$, the free field result is
the larger of the two ratios.

We must emphasize, however, that the AdS/CFT dictionary of section
\ref{dictionary} is by no means complete. Eqs.~\reef{ctfinal} and
\reef{t2holo} establish a connection between the two gravitational
parameters and two CFT couplings characterizing the two- and
three-point functions. However, the GB action \reef{SGB} fixes the form
not just of these CFT correlators but also an infinite series of
$n$-point functions for the stress tensor. Implicitly then, the
parameters controlling the higher $n$-point function are not
independent couplings within the family of CFT's dual to GB gravity or
any simple gravitational action with a finite number of couplings.

Among the interesting results found here were the bounds on the GB
coupling in eq.~\reef{range0}. In section \ref{dictionary}, these
constraints were derived by demanding that the energy flux remained
positive everywhere for certain scattering experiments in the dual CFT.
However, in section \ref{causality}, precisely the same constraints
were reproduced by demanding that causality was respected in the dual
CFT. We should also note that while it did not set a fundamental
constraint, causality violations and negative energy fluxes also appear
in the shear channel at precisely the same critical value of $\lgb$
given in eq.~\reef{lgbshear}. The exact match of the constraints
derived in these two approaches was already observed for GB gravity in
$D=5$ and 7 in \cite{hofman,RobAlex} and \cite{Jan}, respectively. It
was argued in \cite{hofman} that this precise matching is a consequence
of the special two-derivative nature of the GB gravity. While, as
explicitly seen in appendix \ref{misha}, the energy flux calculations
only depend on the three-point function of the stress tensor, in
general, one should expect that the causal propagation of signals in,
\eg a CFT plasma is also effected by higher $n$-point functions. Hence
in general, demanding causal propagation in the CFT provides
independent constraints on the parameters of the dual gravitational
parameters.

Explicitly evaluating the constraints in eq.~\reef{range0} for various
dimensions, we find
\begin{align}
D&=5: \qquad -0.1944 \leq  \lambda_{\text{\tiny{GB}}} \leq 0.0900, \notag \\
D&=6: \qquad -0.2602 \leq  \lambda_{\text{\tiny{GB}}} \leq 0.1523, \notag \\
D&=7: \qquad -0.3125 \leq  \lambda_{\text{\tiny{GB}}} \leq 0.1875, \notag \\
D&=8: \qquad -0.3549 \leq  \lambda_{\text{\tiny{GB}}} \leq 0.2076. \labell{outer}
\end{align}
One can compare these results to the constraints \reef{boundshydro}
arising from demanding subluminal propagation of sound waves in
second-order hydrodynamics. As expected, the above constraints are
always more stringent than those coming from the analysis of
second-order hydrodynamics, as was observed in \cite{RobAlex} for
$D=5$. However, as emphasized there, these two sets of constraints
stand on a completely different footing. If the GB theory lies outside
of the range specified by the causality or energy flux constraints
\reef{range0}, then the gravitational theory and its dual CFT are
fundamentally pathological. In contrast, the constraints
\reef{boundshydro} arising from second-order hydrodynamics only
indicate when a certain approximate description of the CFT plasma
becomes problematic.

One can combine the fundamental constraints \reef{range0} with the
hydrodynamic analysis to examine the behaviour of the limiting
wave-front velocities in the shear \reef{vfrontshear} and sound
\reef{vsound0} channels. In figure \ref{speedbound}, we exhibit the
wave-front speeds in the CFT plasma as a function of $d=D-1$ for GB
gravity with the lower and upper values of $\lgb$ allowed by the bounds
\eqref{range0}, as well as for Einstein gravity (\ie
$\lambda_{\text{\tiny{GB}}}=0$) --- compare to figure 1 of
\cite{Romat2}. Figure \ref{speedbound2} also shows the detailed
behaviour of the wave-front velocity in the sound channel as a function of
$\lgb$ for each dimension. The general trend is that the wave-front
velocities are monotonically decreasing as $d$ grows. Given the
analytic results for Einstein gravity \cite{Romat2}, one sees that in
this case for large values of $d$:
$v_{\text{\tiny{shear}}}^{\text{\tiny{front}}}\sim \sqrt{1/d}$ and
$v_{\text{\tiny{sound}}}^{\text{\tiny{front}}}\sim \sqrt{3/d}$. As seen
in figure \ref{speedbound}, for $d\ge8$, the GB gravity with the
largest (smallest) allowed value of $\lgb$ has the smallest (largest)
wave-front velocities.
\EPSFIGURE{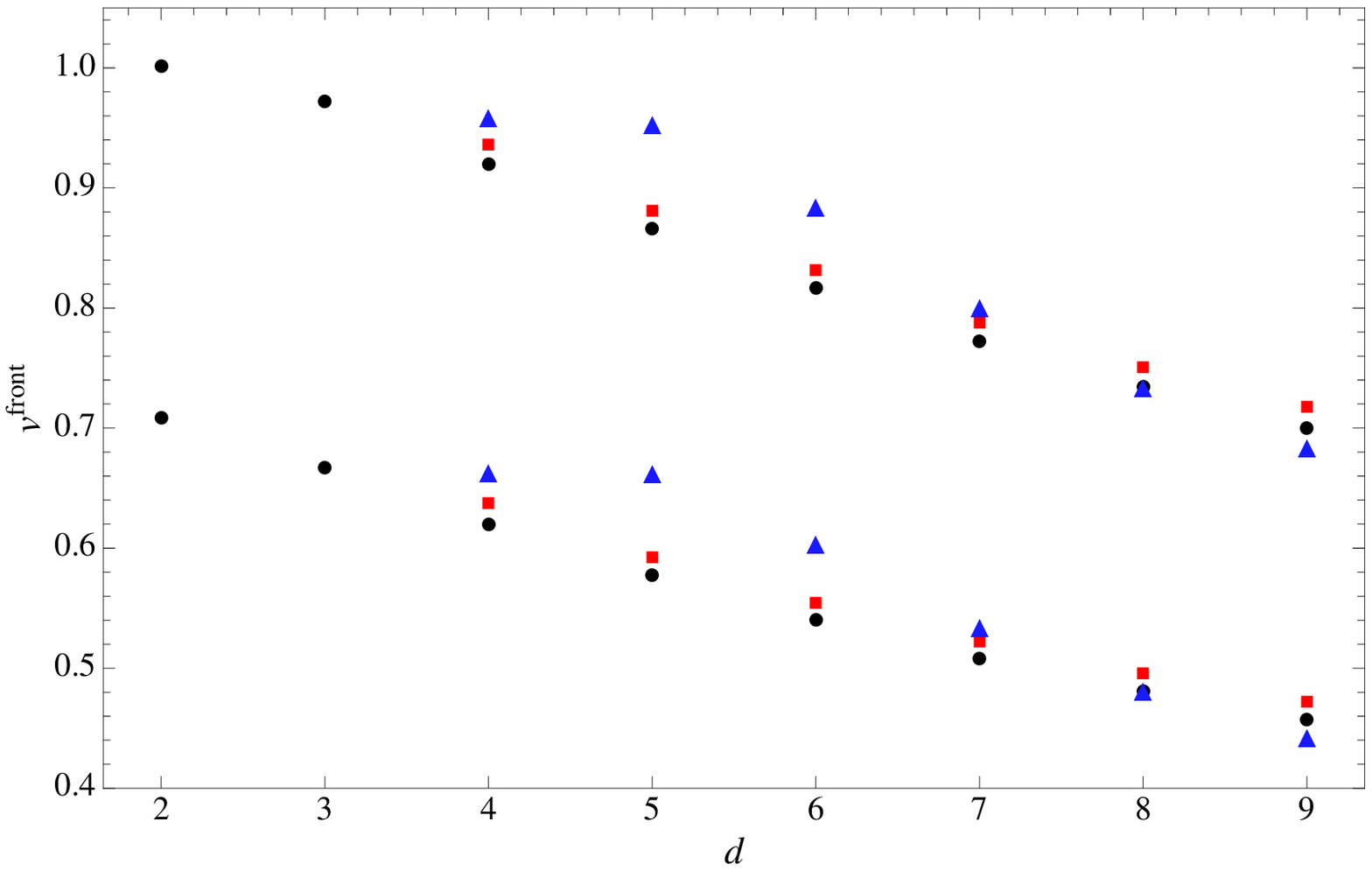,width=12cm} {(Colour online) Limiting
wave-front speeds for the shear (bottom) and sound (top) channel
fluctuations in conformal second-order hydrodynamics.  The black
bullets correspond to the values for Einstein gravity (\ie
$\lambda_{\text{\tiny{GB}}}=0$). The red squares and blue triangles
represent the wave-front velocity for the lower and upper values of
$\lgb$ allowed by the fundamental bounds
\eqref{range0}.\label{speedbound}}
\EPSFIGURE{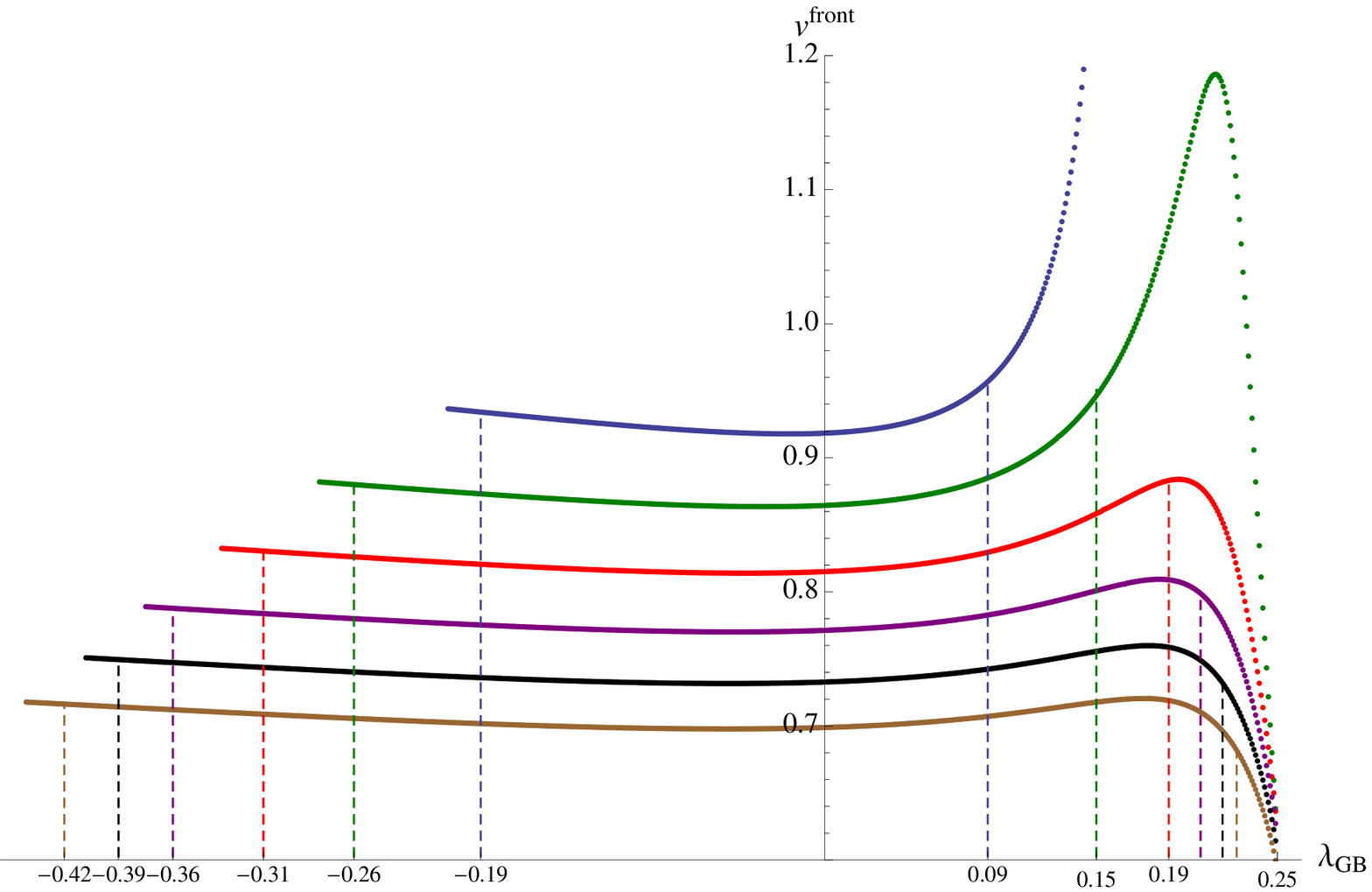,width=13cm} {(Colour online) Wave-front
speed in the sound channel in conformal second-order hydrodynamics in
$d=4,5,6,7,8,9$ (from top to bottom).  The dashed vertical lines
indicate the upper and lower values for $\lambda_{\text{\tiny{GB}}}$
allowed by the bounds \eqref{range0}.  We see that for $d=8,9$, the
speed for the upper value of $\lambda_{\text{\tiny{GB}}}$ is lower than
the Einstein result $\lambda_{\text{\tiny{GB}}}=0$.  This is in
agreement with the behaviour shown in figure \ref{speedbound}.
\label{speedbound2}}

By combining eqs.~\eqref{range0} and \eqref{etas}, one obtains bounds
on the ratio of the the shear viscosity to entropy density in the CFT's
dual to GB gravity. In particular, there is a lower bound which depends
on the dimension of the bulk spacetime as:
\begin{equation}
\frac{\eta}{s} \geq \frac{1}{4\pi} \left[1-\frac{\left(D-1\right)
\left(D-4\right)\left(D^2-3D+8\right)}{2\left(D^2-5D+10\right)^2}\right]\,.
 \labell{lbound}
\end{equation}
For $D=5$, we recover the result $\eta/s \geq 16/25\times(1/4\pi)=
(0.640)/(4\pi)$ originally found in \cite{RobShenker2}. Recently, it
was observed \cite{Jan} that this function \reef{lbound} has a minimum
in the vicinity of $D=9$, as illustrated in figure \ref{plotetas}. More
precisely, one finds a minimum value of $\eta/s\simeq (0.414)/(4\pi)$
for $D\simeq9.207$. Further for large $D$, this lower bound rises again
and asymptotically approaches $\eta/s=1/(8\pi)$. It seems that these
results may provide interesting empirical data in considering the
question of a fundamental lower bound for $\eta/s$, as was conjectured
by \cite{KSS}. However, it is not clear to us what conclusions one may
draw with respect to this question at this point.
\FIGURE{
\includegraphics[width=0.48\textwidth]{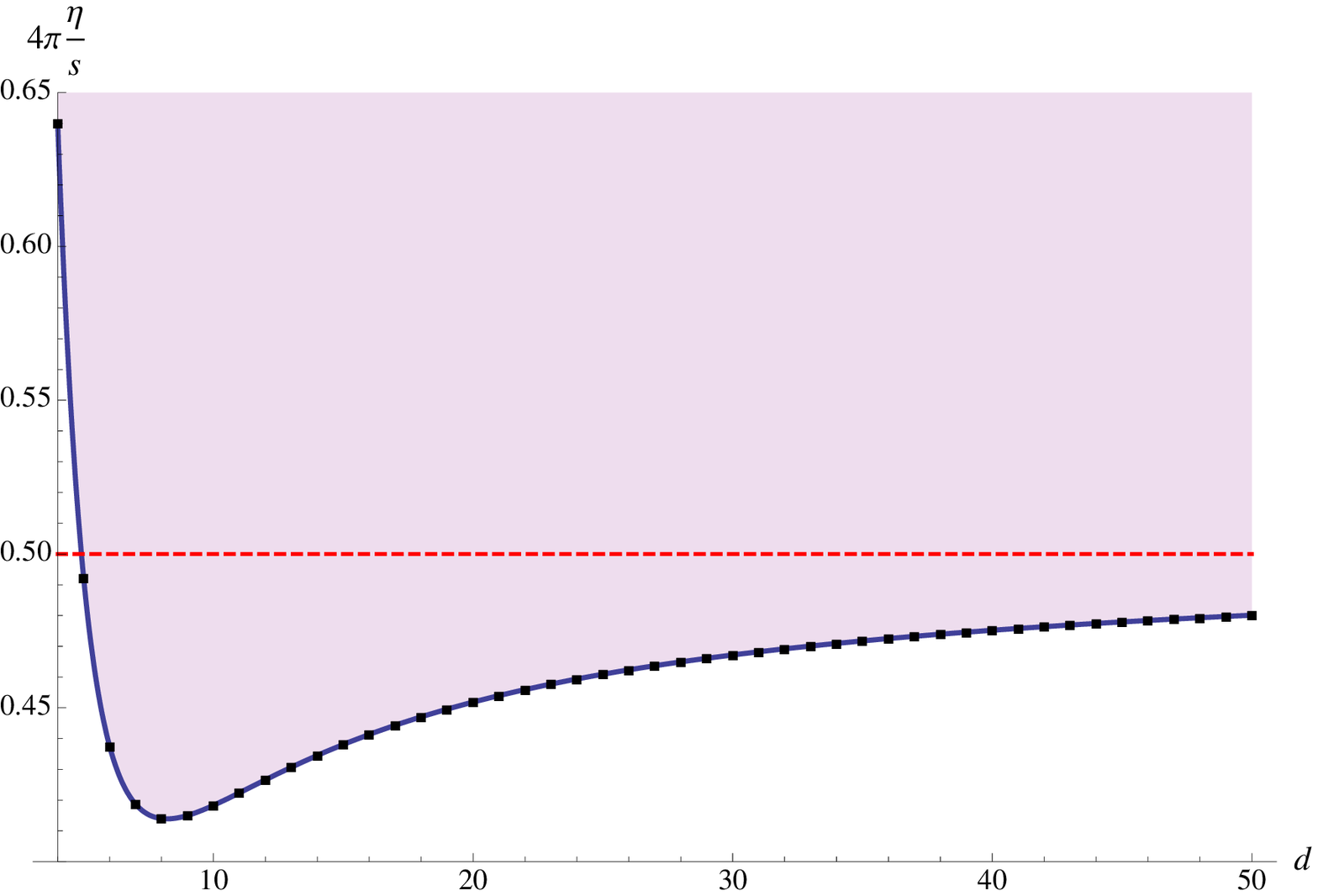}
\includegraphics[width=0.48\textwidth]{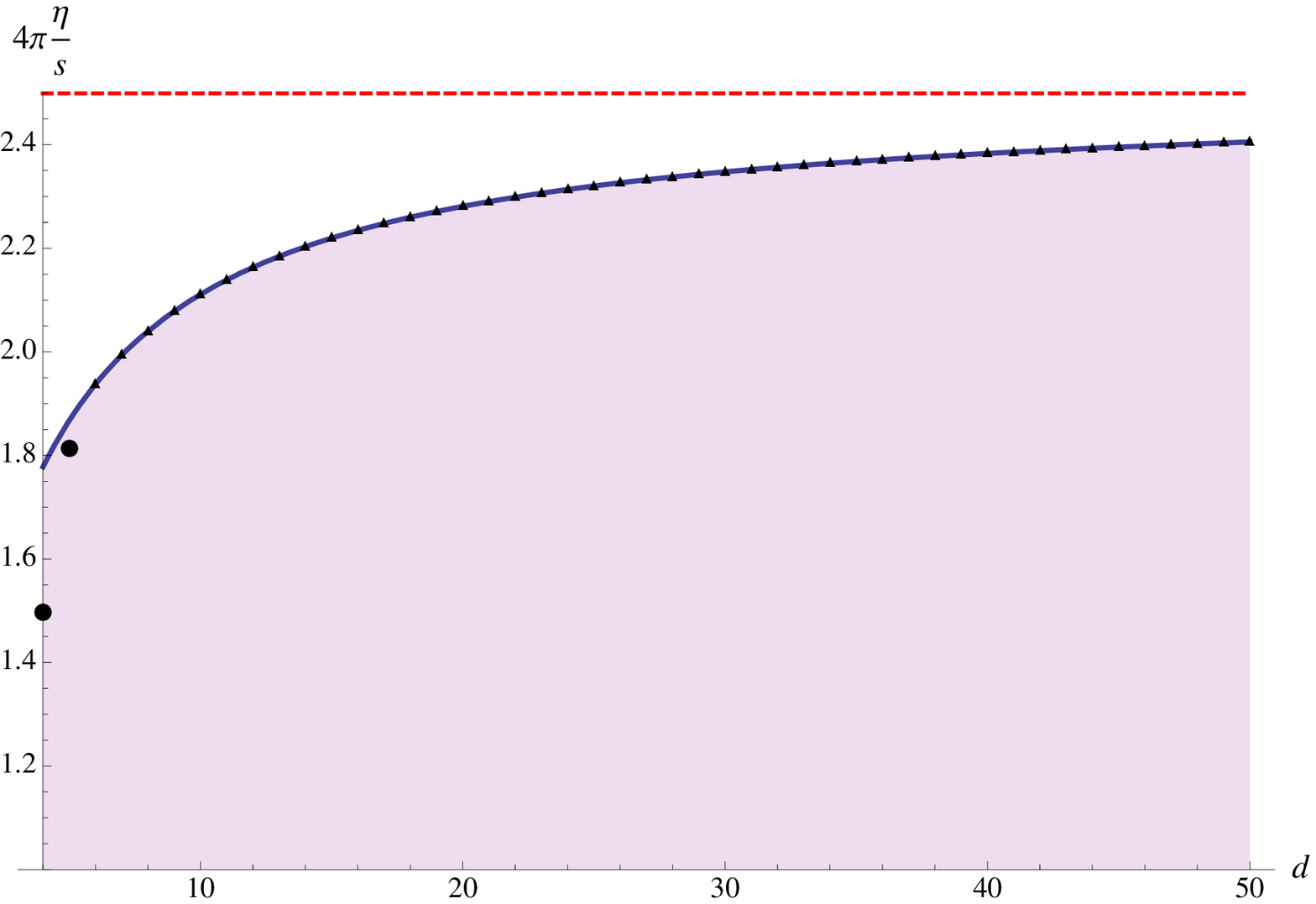}
\caption{(Colour online) On the left, we plot the lower bound
\reef{lbound} on the ratio $\eta/s$ as determined by the upper bound on
$\lgb$ imposed by requiring a causal CFT for various dimensions
$d=D-1$. On the right, we plot the analogous upper bound \reef{upper}
on the ratio $\eta/s$ within the class of CFT's described by GB
gravity. Note that for $d=4$ and 5, the bullets indicate the upper
bound taking into account the plasma instabilities discussed in section
\ref{instab}.} \label{plotetas}}

Within this class of CFT's dual to GB gravity, one also finds an upper
bound for the ratio of the shear viscosity to entropy density
\begin{equation}
\frac{\eta}{s} \leq \frac{1}{4\pi} \left[1+
\frac{(D-1)(3D-1)}{2(D+1)^2}\right]\,. \labell{upper}
\end{equation}
As shown in figure \ref{plotetas}, this upper bound is a monotonically
increasing function which approaches $\eta/s=5/(8\pi)$ asymptotically
for large $D$. Of course, there is no fundamental interpretation of
this upper bound since $\eta/s$ can become arbitrarily large for weakly
coupled theories. Again, this bound only applies for the strongly
coupled CFT's with a holographic description in terms of GB gravity.

As also shown in figure \ref{plotetas}, the previous upper bound is
probably slightly lower in $d=4$ and 5 when one takes into account the
plasma instabilities discussed in section \ref{instab}. These
instabilities do not indicate a fundamental pathology for the GB
theories in the ranges given in eq.~\reef{unstable9}. Rather in these
cases, one simply concludes that a homogenous plasma is an unstable
configuration in the dual CFT. Since the unstable modes have large
momentum, it seems that the plasma wants to ``clump'' into an
inhomogeneous configuration. The immediate implication for our analysis
is that the hydrodynamic calculations, and in particular, the
computation of $\eta/s$, is unreliable in this regime given in
eq.~\reef{unstable9}. From a gravitational perspective, the plasma
instabilities correspond to unstable quasinormal modes of the black
hole background. It would be interesting to carry out a more extensive
analysis of the quasinormal mode spectrum in GB gravity. This would
likely provide greater insight into the nature of these instabilities.

\acknowledgments
 JE thanks Pedro Vieira for useful conversations. RCM would also like
to thank Dam Son, Subir Sachdev and Andrei Starinets for useful
conversations and correspondence. Research at Perimeter Institute is
supported by the Government of Canada through Industry Canada and by
the Province of Ontario through the Ministry of Research \& Innovation.
AB gratefully acknowledges further support by an NSERC Discovery grant
and support through the Early Researcher Award program by the Province
of Ontario. RCM also acknowledges support from an NSERC Discovery grant
and funding from the Canadian Institute for Advanced Research. MFP is
supported by the Portuguese Fundacao para a Ciencia e Tecnologia, grant
SFRH/BD/23438/2005. MFP and MS would also like to thank the Perimeter
Institute for hospitality at various stages of this project.

\appendix

\section{Energy fluxes in terms of the three-point couplings} \label{misha}

In this appendix we present an explicit computation of the energy flux
\reef{basic} for a CFT in an arbitrary number of dimensions. In
particular, $t_2$ and $t_4$ are determined in terms of the coefficients
which appear in the three--point function of the stress--energy tensor
\cite{OP,EO}. Our calculations closely follow those given in
\cite{Jan}.

Let us consider the expression in eq.~(\ref{wok}). Without loss of
generality we can position the detector on the (positive) $x^1$ axis,
\ie $n^i=\delta^i{}_{1}$ as in section \ref{constraints}. We then
introduce the light-cone coordinates $x^{\pm}=t\pm x^1$ and express the
energy flux measured at large distances as
 \begin{equation}
 {\cal E}=\lim_{x^{+}\rightarrow \infty}
 \left({x^{+}-x^{-}\over 2}\right)^{d-2} \int dx^{-}\, T_{--}(x^{+},x^{-})
 \end{equation}
In order to determine $t_2$ and $t_4$, it suffices to compute the
energy one-point function \reef{wok} with specific polarizations which
yield two independent linear combinations of these coefficients.
Recalling the discussion at the beginning of section \ref{constraints},
we will consider polarizations in the tensor and vector channels. For
the tensor channel, we choose a polarization where the only
nonvanishing components are $\epsilon_{\hi\hj}=\epsilon_{\hj\hi}$ where
$\hi$ and $\hj$ are fixed indices with $\hi,\hj>1$ and $\hi\ne\hj$.
Similarly for the vector channel, our polarization will only have
nonvanishing components $\epsilon_{1\hi}=\epsilon_{\hi 1}$ where $\hi$
is again a fixed index with $\hi>1$. The corresponding linear
combinations of $t_2,t_4$ which will appear in eq.~\reef{basic} are
those appearing in the constraints \reef{cstrn1} and \reef{cstrn2}.
That is,
 \beqa
{\rm Tensor:}&&\langle \E(\nvec)\rangle =
\frac{E}{\Omega_{d-2}}\left[1-{1\over d-1} t_2-{2\over
d^2-1}t_4\right]\,,\labell{fluxten}\\
{\rm Vector:}&&\langle \E(\nvec)\rangle = \frac{E}{\Omega_{d-2}}\left[
1+{d-3\over 2(d-1)}t_2-{2\over d^2-1}t_4\right]\,. \labell{fluxvec}
 \eeqa
We will present the details of the calculations for the tensor channel
below. The vector channel calculations are a straightforward extension
of these.

So choosing the tensor polarization above, the numerator of
eq.~(\ref{wok}) is proportional to the following three-point function
\begin{equation}
 f_3(E)\equiv\int d^{\,d} x ~ e^{i E t} \lim_{x_1^{+}\rightarrow \infty}
 \left({x_1^{+}-x_1^{-}\over 2}\right)^{d-2}
 \int dx_1^{-}\ \langle\, T_{\hi\hj}(x)\,T_{--}(x_1\,)\,T_{\hi\hj}(0)
  \,\rangle
 \labell{threepfourier}
\end{equation}
(where again $\hi$ and $\hj$ are fixed indices and so no sum is
intended above). Similarly, the normalisation in the denominator is
provided by the two-point function
\begin{equation}
 f_2(E)\equiv\int d^{\,d} x ~ e^{i E t} \,
 \langle\, T_{\hi\hj}(x)\,T_{\hi\hj}(0)\,\rangle \,.
 \labell{twopfourier}
\end{equation}

In what follows we find it convenient to assume that the total number
of space-time dimensions is even. In particular, this assumption allows
to use the residue theorem in the evaluation of certain integrals.
Since our final result turns out to be insensitive to the parity of the
spacetime dimension, we analytically continue it (and implicitly the
intermediate integrals) to any $d$. Let us proceed with the evaluation
of the numerator \reef{threepfourier} and the denominator
\reef{twopfourier} separately.

We start with the denominator \reef{twopfourier} where we can simply
use the general result for two-point function \reef{twopt}. Inserting
the indices $\hi\hj$ and explicitly evaluating the general expression
yields
\begin{equation}
 \langle T_{\hi\hj}(x)T_{\hi\hj}(0)\rangle={\ct\over 2 \left(x^2\right)^d}
 \left[1-{2\over x^2}\left((x^\hi)^2+(x^\hj)^2\right)+{8\over
 \left(x^2\right)^2 } (x^\hi)^2 (x^\hj)^2\right]
 ~.
\end{equation}
The central charge $\ct$ is given in eq.~(\ref{central}), which we
present here again
\begin{equation}
 \ct={\pi^{{d\over 2}}\over\Gamma\left({d/2}\right)}{(d-1)
 (d+2)\A-2\B-4(d+1)\C \over d(d+2)} \, .
\end{equation}
The constants $\A,\B,\C$ are the parameters which control the
three-point function of the stress tensor \cite{OP,EO}. To evaluate the
integral in eq.~\reef{twopfourier}, we must first provide an
$i\epsilon$ prescription for the operators. Recall that the operators
$T_{\hi\hj}$ in this expression are not time-ordered which then
requires that $t\rightarrow t-i\epsilon$ and so the light-cone
coordinates are both replaced by $x^{\pm}\rightarrow
x^{\pm}-i\epsilon$. The various integrations are then best performed by
first integrating over spatial directions which are perpendicular to
$x^1$, $x^\hi$ and $x^\hj$ using $(d-4)$-dimensional spherical polar
coordinates, subsequently integrating over ($x^\hi$,$x^\hj$)-plane
using polar coordinates and finally the $x^{\pm}$ integrals are
performed using the residue theorem for the poles in these expressions.
The latter integrals are
\begin{equation}
 \int {dx^{+}\over \left(x^{+}-i\epsilon\right)^{{d\over 2}+1}}e^{i \frac{E}{2} x^{+}}
 \int {dx^{-} \over \left(x^{-}-i\epsilon\right)^{{d\over 2}+1}}e^{i \frac{E}{2}x^{-}}
 = {(2\pi i)^2\over \Gamma\left({d+2\over 2}\right)^2} \, \left(i E\over 2\right)^d
 ~.
\end{equation}
This is the essential point where we assume that the dimension $d$ is
even. Otherwise the singularities appearing in these integrals become
branch cuts and the residue theorem can not be applied. Hence, our
final result is given by
\begin{equation}
 f_2(E)= {\pi^{d+2 \over 2}\over (d+1)}~{ C_{T} \over \Gamma(d-1)\Gamma\left({d+2\over 2}\right)}\,
 \left( { E \over 2} \right)^{d}\, .\labell{denom8}
\end{equation}

Next we turn to computing eq.~(\ref{threepfourier}). Again we rely on
the results of \cite{OP,EO} where the form of the three-point function
of the stress tensor was completely fixed using conformal invariance
and energy conservation -- in particular, see (3.15) in
\cite{OP}.\footnote{However, we express our results in terms of the
parameters $\A,\B,\C$ of \cite{EO}. The corresponding parameters in
\cite{OP} are given by: $a=\A/8$, $b=(\B-2\A)/8$ and $c=\C/2$.}
Inserting the appropriate indices and taking the limit
$x_1^{+}\rightarrow\infty$, we find
\begin{equation}
 \lim_{x_1^{+}\rightarrow \infty} \left({x_1^{+}-x_1^{-}\over 2}
 \right)^{d-2}
 \langle\, T_{\hi\hj}(x)\, T_{--}(x_1)\,T_{\hi\hj}(0)\,\rangle = {(-1)^{1-{3\over 2}\,
 d}\,2^{-(d+3)}\,(x^{-})^2 \ t(x)\over
 \left(x^{-}_1-x^{-}+i\epsilon\right)^{d+2\over 2}
 \left(x^{-}_1-i\epsilon\right)^{d+2\over 2} \left(x^2\right)^{d+6\over 2} }
 \label{threeptensor}
\end{equation}
where
\begin{multline}
 t(x)=-4\B\left[8 (x^\hi)^2 (x^\hj)^2-x^2((x^\hi)^2+(x^\hj)^2)\right]
 +4\C\left[16 (x^\hi)^2(x^\hj)^2-2((x^\hi)^2+(x^\hj)^2)x^2+(x^2)^2\right]
 \\
 +\left[ (d+2)(d-2)\A+(d-2)\B-4d\,\C \right]\left[x^2((x^\hi)^2+(x^\hj)^2)-8(x^\hi)^2(x^\hj)^2 \right]
 \\
 +8(d+2)\left((d-2)\,\A+\B-4\C\right)(x^\hi)^2(x^\hj)^2
 +2(d+4)(d-2)(\B-2\C)(x^\hi)^2(x^\hj)^2 ~.
\end{multline}
In the expression above, the $i\epsilon$ prescription was $t_1\to
t_1-i\epsilon$ and $t\to t-2i\epsilon$, giving a larger negative
imaginary part to time in operators standing to the left. Integrating
over $x_1^{-}$ includes only one of the poles in
eq.~(\ref{threeptensor}), irrespective of whether the contour is closed
in the upper or lower half of the $x_1^{-}$-plane. In either case, the
integral yields
\begin{multline}
 \int dx_1^{-}\lim_{x_1^{+}\rightarrow \infty} \left({x_1^{+}-x_1^{-}\over 2}\right)^{d-2}
 \langle\, T_{\hi\hj}(x)\, T_{--}(x_1)\,T_{\hi\hj}(0)\,\rangle
 \\
 = 2\pi i ~ {\Gamma(d+1)\over \Gamma\left( {d+2 \over 2} \right)^2}
 {2^{-(d+3)}\, t(x)\over \left(x^{-}-2i\epsilon\right)^{d-1} \left(x^2\right)^{d+6\over 2}}
 ~.
\end{multline}
Performing the remaining integrals over $x^a$ in
eq.~\reef{threepfourier} as described in the previous case, the final
result becomes
\begin{equation}
 f_3(E)=- \frac{E}{8} \left( { E \over 4} \right)^{d}
 { 2 \pi^{d+4 \over 2} \over \Gamma\left({d+4\over 2}\right)
 \Gamma\left({d+2\over 2}\right)^2}\,
 \left((d-2)\,(d+2)\,\A  + 2\,d\,\B - 4\,d\,\C\right)
 \, .\labell{f3fin}
\end{equation}
Note that in this case, the integrals over the light-cone coordinates
are
\begin{equation}
 \int{dx^{+}\over \left(x^{+}-2i\epsilon\right)^2} e^{i \frac{E}{2} x^{+}}
 \int {dx^{-} \over \left(x^{-}-2i\epsilon\right)^{d+1}} e^{i \frac{E}{2} x^{-}}
 ={\left(2\pi i\right)^2} {1\over \Gamma(d+1)}\left({iE\over2}\right)^{d+1}
 ~,
\end{equation}
and so yield poles irrespective of whether $d$ is even or odd.

Combining the expressions in eqs.~\reef{denom8} and \reef{f3fin} then
yields
\begin{equation}
 {\Omega_{d-2} \over E}~{f_3(E) \over f_2(E)}=-{d+1 \over d}~
 {(d-2)\,(d+2)\,\A  + 2\,d\,\B - 4\,d\,\C\over (d-1)(d+2)\,\A-2\,\B-4(d+1)\,\C}
 ~.\labell{numer8}
\end{equation}
Note that producing this expression requires the use of an identity for
$\Gamma$ functions of the form:
 \beq
\Gamma\!\left(\frac{d}{2}\right)\ \Gamma\!\left(\frac{d-1}{2}\right)
=\sqrt{\pi}\ 2^{2-d}\ \Gamma\left(d-1\right)\,. \labell{curious}
 \eeq
Comparing the previous expression with eq.~\reef{fluxten}, we find our
first linear combination of $t_2$ and $t_4$:
\begin{equation}
 1-{1\over d-1}t_2-{2\over d^2-1}t_4=-{d+1 \over d}~
 {(d-2)\,(d+2)\,\A  + 2\,d\,\B - 4\,d\,\C \over (d-1)(d+2)\,\A-2\,\B-4(d+1)\,\C}
 ~.\labell{tenlin}
\end{equation}
Following the same basic steps as above, one can also compute the flux
with the vector channel polarization to produce the linear combination
of $t_2$ and $t_4$ in eq.~\reef{fluxvec}. The computations are somewhat
more involved in this case and we only present the final result:
\begin{equation}
 1+{d-3\over 2(d-1)}t_2-{2\over d^2-1}t_4
 =(d+1)~{(d-2)\,(d+2)\,\A+(3d-2)\,\B-8\,d\,\C \over (d-1)(d+2)\,\A-2\,\B-4(d+1)\,\C}
 ~.\labell{veclin}
\end{equation}
Combining eqs.~\reef{tenlin} and \reef{veclin}, we can solve for $t_2$
and $t_4$ separately,
\begin{eqnarray}
 t_2&=&\frac{2(d+1)}{d}\,{(d-2)\,(d+2)\,(d+1)\,\A+3\,d^{\,2}\,\B-4\,d\,(2d+1)\,\C
 \over (d-1)(d+2)\,\A-2\,\B-4(d+1)\,\C} ~,
\labell{finalttt} \\ 
t_4&=&
-{(d+1) \over d}~{~(d + 2)\,(2d^2-3d-3)\,\A+2\,d^{\,2}\,(d+2)\,\B-4d\,(d+1)
\,(d+2)\,\C \over (d-1)(d+2)\,\A-2\,\B-4(d+1)\,\C}
 ~.\nonumber
\end{eqnarray}
Note that these expressions for general $d$ reproduce the results
presented in \cite{HM} and \cite{Jan} in $d=4,6$ respectively.

\section{Conformal tensor fields} \label{tensor-app}

In sections \ref{constraints} and \ref{discuss}, we compare various
results for the strongly coupled CFT's dual to GB gravity to those for
massless free fields. As well as conformally coupled scalars and
massless fermions, an ($n$--1)-form potential in $d=2n$ dimensions
yields another free field theory which is also conformally invariant.
The most familiar example is provided by $d=4$, in which case the
potential is just an Abelian vector field. Our final results given in
eq.~\reef{finalout} are general expressions for $\A$, $\B$ and $\C$ for
any value of $n$. The analysis leading to these expression is quite
involved and we only provide some of the salient steps in the
following. Further, for simplicity, we work in a Euclidean-signature
space.

We begin with a free $(n-1)$--form gauge field $A$ in $d=2n$
dimensions. As usual, the corresponding field strength is an $n$-form
given by $F=dA$. With a general background metric, the action for this
system is
\begin{equation}
 I={1\over2(n!)} \int d^{\,2n}x \sqrt{g}g^{\mu_1\nu_1}g^{\mu_2\nu_2}\cdots g^{\mu_n\nu_n}\,
  F_{\mu_1\mu_2\cdots\mu_n}\,F_{\nu_1\nu_2\cdots\nu_n}
 \,, \labell{Fact}
\end{equation}
which is invariant under conformal rescalings, $g_{\mu\nu}\to
e^{2\phi}g_{\mu\nu}$. To simplify the subsequent calculations, we now
fix the background metric to be simply flat space, \ie $g_{\mu\nu}=
\delta_{\mu\nu}$. The corresponding energy momentum tensor is given by
\begin{equation}
 T_{\mu\nu}={1\over (n-1)!} F_{\mu \, \alpha_2\cdot\cdot\cdot\alpha_n}F_{\nu}^{~\alpha_2\cdot\cdot\cdot\alpha_n}
 -{1\over2(n!)}\delta_{\mu\nu}\,F_{\alpha_1\cdot\cdot\cdot\alpha_n}F^{\alpha_1\cdot\cdot\cdot\alpha_n}
 \quad .
 \label{EMT}
\end{equation}

Because the theory is invariant under gauge transformations $\delta
A=d\Lambda$, in principle, we should introduce gauge-fixing terms to
the action \reef{Fact}. However, correlation functions involving only
gauge invariant operators will be unaffected by the choice of
gauge-fixing and any ghost fields and so we do not consider the
complete details here. With the natural generalization of Feynman
gauge, the two-point function of the potential becomes
\begin{equation}
 \langle \, A_{\mu_1\cdot\cdot\cdot\mu_{n-1}}(x_1) A_{\nu_1\cdot\cdot\cdot\nu_{n-1}}(x_2)\,\rangle
 ={\Gamma(n-1) \over 4\,\pi^n}{1 \over (x_{12}^{2})^{n-1}}
 \sum_{\sigma\in S_{n-1}}\mathrm{sign}(\sigma)\,\delta_{\mu_1\nu_{\sigma(1)}}
 \cdot\cdot\cdot\delta_{\mu_{n-1}\nu_{\sigma(n-1)}}
\end{equation}
where $S_{n-1}$ is a permutation group of $n-1$ elements $1,2,..,n-1$.
Differentiating the above two-point function with respect to
$x_1^{\mu_n}$ and $x_2^{\nu_n}$ and subsequently antisymmetrizing the
result with respect to $\mu$'s and $\nu$'s yields the following basic
gauge invariant two-point function of the field strength
\begin{equation}
 \langle \, F_{\mu_1\cdot\cdot\cdot\mu_n}(x) F_{\nu_1\cdot\cdot\cdot\nu_n}(0)\,\rangle
 ={n \over \Omega_{2n-1}}~{J_{\mu_1\cdot\cdot\cdot\mu_n\,;\,\nu_1\cdot\cdot\cdot\nu_n} \over (x^{2})^n}
 \quad ,
 \label{basic-2p-func}
\end{equation}
where
\begin{equation}
 J_{\mu_1\cdot\cdot\cdot\mu_n\,;\,\nu_1\cdot\cdot\cdot\nu_n}(x)=
 \sum_{\sigma\in S_n}\mathrm{sign}(\sigma)\,I_{\mu_1\nu_{\sigma(1)}}(x)\cdot\cdot\cdot I_{\mu_{n}\nu_{\sigma(n)}}(x)
 \quad .
 \label{defJ}
\end{equation}
As in eq.~\reef{tensor}, we have defined the orthogonal matrix
 \beq
I_{\mu\nu}(x)=\delta_{\mu\nu}-2\frac{x_\mu\,x_\nu}{x^2}\,.
 \labell{tensor9}
 \eeq
After a lengthy calculation, we arrive at the connected two-point
function of the energy momentum tensor
 \beq
 \langle T_{\mu\nu}(x)\,T_{\alpha\beta}(0)\rangle=
 {n^2\,\Gamma[2n-1] \over 2 \pi^{2n}}\,{1 \over (x^2)^{2n}}\,
 \mathcal{I}_{\mu\nu,\alpha\beta}(x)
 ~,\labell{toostressed}
 \eeq
where $\mathcal{I}_{\mu\nu,\alpha\beta}(x)$ is defined as in
eq.~\reef{tensor0}. Of course, our result matches the expected form
given in eq.~\reef{twopt} where
\begin{equation}
 C_{\,T}={d^{\,2} \over 8 \, \pi^{\,d}}\,\Gamma[d-1] \quad .
 \label{Centcharge}
\end{equation}

Proceeding further requires calculating the three-point function of the
stress tensor. While such a computation is straightforward, it would be
extremely tedious to carry out in general. The calculations are greatly
simplified by going to the collinear frame \cite{OP}, for which the
three points are chosen to lie on a straight line. In this case, the
form of the three-point function is given by eqs.~(4.20) and (4.21) of
\cite{OP}. Since we wish to fix the values of $\mathcal{A}$,
$\mathcal{B}$ and $\mathcal{C}$, it is enough to supplement
eq.~(\ref{Centcharge}) with two extra relations between these
constants. We choose to compute coefficients $\alpha$ and $\gamma$
defined in eq.~(4.21) of \cite{OP}.

For simplicity, we exploit the rotational and translational symmetry of
the problem to choose the following three points for the insertions of
the energy momentum tensor: $x_1=(x,\mathbf{0})$, $x_2=(y,\mathbf{0})$,
$x_3=0$. As a result, $I_{\mu\nu}$ becomes a constant orthogonal matrix
given by
\begin{equation}
 I_{\mu\nu}=\delta_{\mu\nu}-2\,\delta_{1\mu}\delta_{1\nu}~.
 \label{I-collinear}
\end{equation}
With some work, we find
\begin{equation}
 \langle T_{11}(x)T_{11}(y)T_{11}(0) \rangle =
 -{n^2 \, \Gamma(n)\Gamma(2n) \over 4 \, \pi^{3n}}
 { 1 \over x^{2n}y^{2n}(x-y)^{2n}}
 \quad ,
\end{equation}
and hence, in the notation of \cite{OP},
\begin{equation}
 \alpha=-{d^2 \, \Gamma(d/2)\Gamma(d) \over 16 \, \pi^{3d/2}}
 \quad .
 \label{alpha}
\end{equation}
Similarly, we obtain
\begin{equation}
 \langle T_{i1}(x)T_{j1}(y)T_{11}(0) \rangle = \, {n^3 \, \Gamma(n) \, \Gamma(2n-1)\over 4 \pi^{3n}}
 { \delta_{ij} \over x^{2n}y^{2n}(x-y)^{2n}}
 \quad ,
\end{equation}
which yields
\begin{equation}
 \gamma=\, {d^3 \, \Gamma(d/2) \, \Gamma(d-1)\over 32 \, \pi^{\,3d/2}}
 \quad .
 \label{gamma}
\end{equation}
According to eqs.~(4.22$\,-\,$4.25) of \cite{OP},
\begin{eqnarray}
 \alpha&=&-{(d-1)(d^{\,3}-2\,d^{\,2}-d+4) \over 2\,d^{\,2}}\,\mathcal{A}+{(d-1)^3\over 4\,d}\,\mathcal{B}
 -{(d+1)(d-1)^2(d-4) \over 2\,d^{\,2}}\,\mathcal{C}~,
 \nonumber\\
 \gamma&=&{d^{\,3}-\,d^{\,2}-2\,d+4 \over 8\,d}\,\mathcal{A}+{(d-3)\over 8}\,\mathcal{B}
 -{(d+1)(d-1) \over 2\,d}\,\mathcal{C}~.
\end{eqnarray}
Hence combining these two equations with
(\ref{central}),(\ref{Centcharge}),(\ref{alpha}),(\ref{gamma}) and
solving for $\mathcal{A}$, $\mathcal{B}$ and $\mathcal{C}$ yields
\begin{eqnarray}
 \mathcal{A}&=&-{d^3 \Gamma(d-1)\Gamma(d/2) \over 8(d-3)\pi^{3d/2}}
 ~,
 \nonumber\\
 \mathcal{B}&=&(d-2)\,\mathcal{A}
 ~,
 \labell{finalout}\\
 \mathcal{C}&=&{d-2 \over 2}\,\mathcal{A}
 ~. \nonumber
\end{eqnarray}
These general results agree with those derived for $d=4$ in \cite{OP}
and for $d=6$ in \cite{Jan}.

\section{Useful expressions} \label{appA}
In order to write our equations as neatly as possible, we define the
following expressions, which appear repeatedly in our analysis:
\begin{equation*}
\mathcal{K}=u^{\frac{D-1}{2}}=1-f+\lambda_{\text{\tiny{GB}}}f^2,
\label{K}
\end{equation*}
\begin{equation*}
\mathcal{M}=1-2\frac{(D-1)}{(D-3)}\lambda_{\text{\tiny{GB}}} -2\frac{(D-5)}{(D-3)}\lambda_{\text{\tiny{GB}}} f \left(1-\lambda_{\text{\tiny{GB}}}f \right),
\end{equation*}
\begin{equation*}
\mathcal{N}=
\left[1-2\lambda_{\text{\tiny{GB}}} f \right]^2,
\label{N}
\end{equation*}
\begin{align*}
\mathcal{P}=&-\frac{1}{2} (D-1)\left[1-\frac{2(D-1)}{(D-3)}\lambda_{\text{\tiny{GB}}}\right] + \left[1+\frac{2(D-1)(D-6)}{(D-3)}\lambda_{\text{\tiny{GB}}}\right] f- \notag \\
& +\frac{1}{2(D-3)(D-4)} \left[172-91D+16D^2-D^3 + 8(D-1)(D-4)(2D-9) \lambda_{\text{\tiny{GB}}} \right] \lambda_{\text{\tiny{GB}}} f^2 + \notag \\
& +\frac{2}{(D-3)^2} \left[87 -44D +5D^2+4(D-1)(D-3)(D-4) \lambda_{\text{\tiny{GB}}} \right] \lambda_{\text{\tiny{GB}}}^2 f^3 - \notag \\
& +\frac{(D-5)(D+15)}{(D-3)} \lambda_{\text{\tiny{GB}}}^3 f^4 +\frac{8(D-5)}{(D-3)}\lambda_{\text{\tiny{GB}}}^4 f^5,
\end{align*}
\begin{align*}
\mathcal{R}=\frac{(D-1)^2}{64\fin} \Biggl[&4\fin\textswab{w}^2 \left[1-\frac{2(D-1)}{(D-3)}\lambda_{\text{\tiny{GB}}}
\right]\notag\\
& + 4 \biggl[ \textswab{q}^2 \left[\frac{2(D-1)}{(D-3)(D-4)}\lambda_{\text{\tiny{GB}}} \left[2(D-1)\lambda_{\text{\tiny{GB}}}+D-6\right]-1\right]+ \notag \\
& +\frac{2 \fin\textswab{w}^2}{(D-3)} \lambda_{\text{\tiny{GB}}} \left[4(D-1)\lambda_{\text{\tiny{GB}}} -3D+11\right] \biggr] f -\notag \\
& +\frac{4}{(D-3)(D-4)}\biggl[2\textswab{q}^2 \left[4(D-1)(2D-7)\lambda_{\text{\tiny{GB}}}-3D(D-7)-42\right] +\notag \\
&+2(D-4) \fin\textswab{w}^2 \lambda_{\text{\tiny{GB}}} \left[4(D-1)\lambda_{\text{\tiny{GB}}} -7D+31\right] \biggr]\lambda_{\text{\tiny{GB}}} f^2 +\notag \\
& +\frac{8}{(D-3)(D-4)} \biggl[4\lambda_{\text{\tiny{GB}}}  \left[ \textswab{q}^2(D-1)(2D-7)-2(D-4)(D-5)\fin \textswab{w}^2 \right]- \notag \\
& + \textswab{q}^2\left[5D(D-9)+112\right] \biggr] \lambda_{\text{\tiny{GB}}}^2 f^3 + \frac{16(D-5)}{(D-3)(D-4)} \bigl[2\textswab{q}^2(D-7)+ \notag \\
&+2(D-4) \fin \textswab{w}^2\lambda_{\text{\tiny{GB}}} \bigr]\lambda_{\text{\tiny{GB}}}^3 f^4 -\frac{16(D-5)(D-7) \textswab{q}^2}{(D-3)(D-4)} \lambda_{\text{\tiny{GB}}}^4 f^5 \Biggr],
\end{align*}
\begin{equation*}
\mathcal{S}= -\frac{3}{4}(D-1) + \frac{1}{4} \left[3+D+10(D-1)\lambda_{\text{\tiny{GB}}}\right]f - \frac{1}{4}(5D+11)\lambda_{\text{\tiny{GB}}}f^2
+ \frac{1}{2}(D+7)\lambda_{\text{\tiny{GB}}}^2 f^3,
\end{equation*}
\begin{equation*}
\mathcal{T}= -\frac{(D-1)}{4} + \frac{1}{4} \left[5-D+6(D-1)\lambda_{\text{\tiny{GB}}}\right]f + \frac{1}{4}(D-17)\lambda_{\text{\tiny{GB}}}f^2
- \frac{1}{2}(D-9)\lambda_{\text{\tiny{GB}}}^2 f^3,
\end{equation*}
\begin{equation*}
\mathcal{V}= \frac{1}{2(D-2)} \biggl[ (D-1) + \left[D-3-6(D-1)\lambda_{\text{\tiny{GB}}}\right]f - (D-9)\lambda_{\text{\tiny{GB}}}f^2 + 2(D-5)\lambda_{\text{\tiny{GB}}}^2 f^3 \biggr].
\end{equation*}

\section{Coefficients for the sound channel} \label{appB}
\begin{align*}
\text{Numerator  of }& \mathcal{C}^{(1)}_{\text{\tiny{sound}}} = \, 2(D-3) \textswab{q}^2 f^2 \mathcal{M} \biggl[ 2(D-1)(D-2) \lambda_{\text{\tiny{GB}}} \fin\textswab{w}^2 \mathcal{M} \mathcal{N} \mathcal{K} + \textswab{q}^2 \mathcal{N} \mathcal{P}+ \notag \\
&-(D-3) \textswab{q}^2 \mathcal{M}^2 \mathcal{T} + \mathcal{M} \Bigl[(D-3) \left( \textswab{q}^2 \mathcal{P} - (D-2)\fin\textswab{w}^2 \mathcal{N}^2 \right) -  \textswab{q}^2 \mathcal{N} \mathcal{T} \Bigr] \biggr] +\notag \\
&- (D-2) \mathcal{N}^{1/2} \mathcal{P} \Bigl[(D-1) \textswab{q}^2 \mathcal{K} -2(D-2)\fin\textswab{w}^2 \mathcal{N}^{1/2} \Bigr] \Bigl[ \textswab{q}^2 \mathcal{V}-\fin\textswab{w}^2\mathcal{N}\Bigr]  + \notag \\
&- \textswab{q}^2 f \Biggl[-2(D-1)^2(D-2) \lambda_{\text{\tiny{GB}}} \mathcal{M} \mathcal{N}^{1/2} \mathcal{K}^2 \left(\fin\textswab{w}^2 \mathcal{N} - \textswab{q}^2 \mathcal{V} \right) + \notag \\
&+ (D-1)(D-3) \mathcal{M} \mathcal{N}^{1/2} \mathcal{K} \Bigl[ (D-2)\fin\textswab{w}^2 \mathcal{N}^2 +  \textswab{q}^2  \left( \mathcal{M} \mathcal{T} - \mathcal{P}\right) + \notag \\
&+(D-2) \textswab{q}^2 \left(2 \mathcal{M} -\mathcal{N}\right) \mathcal{V} \Bigr]+ 2(D-2) \Bigl[\fin\textswab{w}^2 \left((D-3)\mathcal{M} - \mathcal{N}\right) \mathcal{N} \mathcal{P} + \notag \\
&+  \textswab{q}^2 \bigl[\mathcal{N} \mathcal{P} +(D-3)\mathcal{M}^2\mathcal{S} \bigr] \mathcal{V} \Bigr] \Biggr],
\end{align*}
\begin{align*}
\text{Denominator of }\mathcal{C}^{(1)}_{\text{\tiny{sound}}}= &\, (D-2) u f \mathcal{M}\mathcal{N} \left(\fin\textswab{w}^2 \mathcal{N} - \textswab{q}^2 \mathcal{V} \right) \biggl[2\textswab{q}^2 f \left((D-3)\mathcal{M}+\mathcal{N}\right) + \notag \\
&(D-1) \textswab{q}^2 \mathcal{N}^{1/2} \mathcal{K} - 2(D-2)\fin\textswab{w}^2 \mathcal{N}\biggr],
\end{align*}

\text{Numerator of } $\mathcal{C}^{(2)}_{\text{\tiny{sound}}}$ =
\begin{align*}
&(D-1) \Biggl[ \mathcal{N}^{1/2} \left[(D-1) \textswab{q}^2 \mathcal{K} -2(D-2)\fin\textswab{w}^2 \mathcal{N}^{1/2}\right] \biggl[ \textswab{q}^2 \mathcal{K} \Bigl[-(D-1)^2 u \mathcal{N} \Bigl[-2 \textswab{q}^2 \mathcal{P}+ \notag \\
&+\mathcal{M} \left(5(D-2)\fin \textswab{w}^2 \mathcal{N} +2 \textswab{q}^2 \mathcal{T}\right) \Bigr] + (D-2) \left[\mathcal{M}\left((D-1)^2  \textswab{q}^2 u \mathcal{N} -8(D-2)\fin\mathcal{S}\right)-8(D-2)\fin\mathcal{P}\right] \mathcal{V} \Bigr]+ \notag \\
&+ 4(D-1)(D-2) \fin \textswab{q}^2 \mathcal{N}^{1/2} \mathcal{K}^2 \left[\mathcal{P}-\mathcal{M} \left(\mathcal{T}+2(D-2)\mathcal{V}\right)\right] + 2(D-1)(D-2)\fin u \textswab{w}^2 \mathcal{N}^{3/2} \times \notag \\
& \left[-2 \textswab{q}^2 \mathcal{P}+ \mathcal{M} \left[(D-2)\fin\textswab{w}^2 \mathcal{N} -  \textswab{q}^2 \left(2\mathcal{S}+(D-2)\mathcal{V}\right)\right]\right] \biggr] + 4 \textswab{q}^2 f^2 \biggl[ 8(D-1)(D-2)^2\fin \lambda_{\text{\tiny{GB}}}  \mathcal{M} \mathcal{K}^2 \times \notag \\
&\Bigl[(D-3)\fin\textswab{w}^2 \mathcal{M} \mathcal{N} +  \textswab{q}^2 \left((D-3)\mathcal{M} + \mathcal{N}\right)\mathcal{V} \Bigr] + (D-1) \textswab{q}^2 u \mathcal{N}^{1/2} \left((D-3)\mathcal{M} + \mathcal{N}\right) \times \notag \\
&\Bigl[ 2 \textswab{q}^2 \mathcal{N} \mathcal{P} - (D-3)\mathcal{M}^2 \left[(D-2)\fin\textswab{w}^2 \mathcal{N} + \textswab{q}^2 \left(2\mathcal{T}-(D-2)\mathcal{V}\right)\right] + \mathcal{M} \bigl[-(D-2)(2D-5)\fin\textswab{w}^2 \mathcal{N}^2 + \notag \\
&+2(D-3) \textswab{q}^2 \mathcal{P} +  \textswab{q}^2 \mathcal{N} \left((D-2)\mathcal{V}-2\mathcal{T}\right) \bigr] \Bigr]+2(D-2)\fin\mathcal{K} \Bigl[2 \textswab{q}^2 \mathcal{N}^2 \mathcal{P} - 4(D-3)^2  \textswab{q}^2 \mathcal{M}^3 \mathcal{T} + \notag \\
&+ \textswab{q}^2 \mathcal{M} \mathcal{N} \bigl[ 2(D-1)^2 \lambda_{\text{\tiny{GB}}} u \textswab{w}^2 \mathcal{N}^{3/2} + 6(D-3) \mathcal{P} - 2 \mathcal{N} \left(\mathcal{T}+(D-2)(D-3)\mathcal{V}\right) \bigr] +\notag \\
&+ (D-3) \mathcal{M}^2 \Bigl[ 2(D-1)^2\lambda_{\text{\tiny{GB}}} \textswab{q}^2 u \textswab{w}^2 \mathcal{N}^{3/2} -2(D-2)(D-3) \fin \textswab{w}^2 \mathcal{N}^2 + 4(D-3)  \textswab{q}^2 \mathcal{P} +\notag \\
&- 2 \textswab{q}^2 \mathcal{N} \left(3\mathcal{T} +(D-2)(D-3)\mathcal{V} \right) \Bigr] \Bigr] \biggr]
+ 4 \textswab{q}^2 f \biggl[4(D-1)^2(D-2)^2\fin^2 \lambda_{\text{\tiny{GB}}}  \textswab{w}^2 \mathcal{M} \mathcal{N}^{3/2} \mathcal{K}^3 + \notag \\
&+2(D-1)(D-2)\fin u \textswab{w}^2 \mathcal{N}^{3/2} \Bigl[-2 \textswab{q}^2 \mathcal{N} \mathcal{P} +(D-3) \mathcal{M}^2 \left[(D-2)\fin\textswab{w}^2 \mathcal{N} +  \textswab{q}^2 \left(\mathcal{T}-\mathcal{S}-(D-2)\mathcal{V}\right)\right] + \notag \\
&+\mathcal{M} \bigl[(D-2)^2\fin \textswab{w}^2 \mathcal{N}^2 -2(D-3) \textswab{q}^2 \mathcal{P} +  \textswab{q}^2 \mathcal{N} \left(\mathcal{T}-\mathcal{S}-(D-2)\mathcal{V}\right)\bigr] \Bigr] + (D-1)(D-2)\fin \mathcal{N}^{1/2} \mathcal{K}^2 \times \notag \\
&\Bigl[4  \textswab{q}^2 \mathcal{N} \mathcal{P} -2(D-3) \textswab{q}^2 \mathcal{M}^2 \left(3\mathcal{T} + 4(D-2)\mathcal{V}\right) + \mathcal{M} \bigl[2(D-1)^2 \lambda_{\text{\tiny{GB}}} \textswab{q}^2 u \textswab{w}^2 \mathcal{N}^{3/2} + \notag \\
&-2(D-2)(D-3)\fin\textswab{w}^2 \mathcal{N}^2 + 6(D-3) \textswab{q}^2 \mathcal{P} - 8(D-2)^2\fin \lambda_{\text{\tiny{GB}}} \textswab{w}^2 \mathcal{N}^{1/2} \mathcal{V} - 4 \textswab{q}^2 \mathcal{N} \left(\mathcal{T}+(D-2)\mathcal{V}\right) \bigr] \Bigr] + \notag \\
&+\mathcal{K} \Bigl[ 2\mathcal{N} \mathcal{P} \left[\left((D-1)^2 \textswab{q}^4 u -2(D-2)^2\fin^2 \textswab{w}^2\right)\mathcal{N} -2(D-2)^2 \fin \textswab{q}^2 \mathcal{V}\right] + (D-3)\mathcal{M}^2 \times \notag\\
&\bigl[ \mathcal{N} \bigl[-3(D-1)^2(D-2)\fin \textswab{q}^2 u \textswab{w}^2 \mathcal{N} - 2 \left((D-1)^2  \textswab{q}^4 u -2(D-2)^2\fin^2 \textswab{w}^2\right) \mathcal{T} \bigr] +\notag\\
&+ (D-2)  \textswab{q}^2  \left((D-1)^2  \textswab{q}^2 u \mathcal{N} -8 (D-2)\fin\mathcal{S}\right) \mathcal{V} \bigr] + \mathcal{M} \mathcal{N} \bigl[-4(D-1)^2(D-2)^2\fin^2 \lambda_{\text{\tiny{GB}}} u \textswab{w}^4 \mathcal{N}^{3/2} +\notag\\
&-D(D-1)^2 (D-2) \fin \textswab{q}^2 u \textswab{w}^2 \mathcal{N}^2 + 2(D-3) \bigl((D-1)^2 \textswab{q}^4 u - 6(D-2)^2\fin^2 \textswab{w}^2 \bigr) \mathcal{P} - 4(D-2)^2 \fin \textswab{q}^2  \mathcal{S} \mathcal{V} + \notag \\
&+\mathcal{N} \bigl[-2\mathcal{T} \left((D-1)^2 \textswab{q}^4 u -2 (D-2)^2\fin^2 \textswab{w}^2\right)+(D-2)\bigl((D-1)^2  \textswab{q}^4 u + 4(D-2)^2(D-3)\fin^2\textswab{w}^2\bigr)\mathcal{V}\bigr] \bigr] \Bigr] \biggr] \Biggr],
\end{align*}
\begin{align*}
\text{Denominator of }\mathcal{C}^{(2)}_{\text{\tiny{sound}}}=  \, &32(D-2)^2\fin u^2 f^2 \mathcal{M}\mathcal{N}^{3/2} \left(\fin\textswab{w}^2 \mathcal{N} - \textswab{q}^2 \mathcal{V} \right) \biggl[2\textswab{q}^2 f \left((D-3)\mathcal{M}+\mathcal{N}\right) \notag \\
&+ (D-1) \textswab{q}^2 \mathcal{N}^{1/2} \mathcal{K} - 2(D-2)\fin\textswab{w}^2 \mathcal{N} \biggr],
\end{align*}

\end{document}